\documentclass[11pt,a4paper]{article}

\usepackage[utf8]{inputenc}
\usepackage[T1]{fontenc}
\usepackage{lmodern}
\usepackage[margin=1in]{geometry}
\usepackage{microtype}
\usepackage{amsmath,amssymb}
\usepackage{booktabs}
\usepackage{graphicx}
\usepackage{xcolor}
\usepackage[numbers,sort&compress]{natbib}
\usepackage{hyperref}
\usepackage{enumitem}
\usepackage[font=small,labelfont=bf]{caption}
\usepackage{subcaption}
\usepackage{placeins}
\usepackage{indentfirst}
\usepackage[normalem]{ulem}

\hypersetup{
  colorlinks=false,
  pdfborder={0 0 1},
  linkbordercolor={1 0 0},
  citebordercolor={0 1 0},
  urlbordercolor={0 0 1}
}

\newcommand{\E}{\mathbb{E}}
\newcommand{\Prob}{\mathbb{P}}
\newcommand{\dd}{\,\mathrm{d}}

\newcommand\blfootnote[1]{%
  \begingroup
  \renewcommand\thefootnote{}\footnote{#1}%
  \addtocounter{footnote}{-1}%
  \endgroup
}

\title{Deep Learning of Robust Market Making \\ under Regime-Switching Order Flow}
\author{%
  Felipe Moret \quad Fabrizio Lillo \\[2pt]
  \small Scuola Normale Superiore di Pisa \\
  \small \texttt{felipebueno.moret@sns.it}, \;
  \texttt{fabrizio.lillo@sns.it}%
}
\date{}

\graphicspath{{figures/}{./}}

\begin{document}
\maketitle
\blfootnote{\raggedright\urlstyle{same} Code, trained models, and paper source:
\url{https://github.com/felipemoret77/robust-deep-market-making}.}

\begin{abstract}
Classical market-making strategies based on stochastic control, such as
the Avellaneda--Stoikov and the Gu\'{e}ant--Lehalle--Fernandez-Tapia (GLFT)
extension, provide closed-form quoting approximations, but rest on assumptions that
break down at realistic microstructure timescales.  One of them is that
order flow is stationary, while empirical evidence points to the existence
of regimes, possibly associated with algorithmic execution of metaorders.
In our experiments, persistent directional flow makes stationary
controllers unprofitable.  In this paper, we
develop a deep reinforcement-learning market maker (RLMM)---a Rainbow-style
distributional deep Q-network which is calibrated and tested in a
zero-intelligence limit order book.  We find that, in the stationary
setting, RLMM outperforms GLFT at comparable levels of profit and loss
(PnL) dispersion.
The stationarily trained RLMM is more robust to flow asymmetry
than GLFT, but still suffers inventory saturation and large drawdowns
under persistent directional flow.  Training a separate RLMM under
regime-switching flow, with a Bayesian flow-belief signal and a
queue-adjusted quote-exposure signal, restores positive mean PnL in our
experiments.  A final scenario-bandit step prioritizes low-return
scenarios and improves performance under varying
regime persistence and directional correlation.
\end{abstract}

\section{Introduction}

Market makers (MMs) are the intermediaries who keep modern electronic
markets liquid: they continuously post two-sided quotes, standing
ready to both buy and sell, and earn the bid--ask spread as
compensation for supplying immediacy to the rest of the market.  This
service is not riskless.  Every fill changes the market maker's
inventory, and an intermediary who accumulates a large directional
position is exposed to adverse price moves before that position can
be unwound.  The intensity and directional balance of order flow also vary over
time. Liquidity demand arrives in bursts, buy or sell pressure can
persist for long periods, and the prevailing regime can change.
Robust quoting must therefore contend not merely with average
conditions but with the persistent, regime-dependent flow that
dominates at microstructure timescales---a thread we develop across
the paper's staged training pipeline.

Market making is naturally framed as a sequential stochastic-control
problem: the market maker chooses bid and ask quotes, the market
responds with random order arrivals and cancellations, and the market
maker must trade off spread revenue against inventory risk.  The
canonical analytical benchmarks are Avellaneda--Stoikov
(AS)~\citep{avellaneda2008} and the
Gu\'{e}ant--Lehalle--Fernandez-Tapia (GLFT)
extension~\citep{gueant2013dealing,gueant2017optimal}, both of which
yield closed-form quoting approximations under stylized assumptions: a
diffusive mid-price, exponentially decaying fill intensity in quote
distance, and stationary, symmetric order flow; see
\citet{cartea2015algorithmic} and \citet{gueant2016financial} for
textbook treatments of optimal market making, and \citet{cartea2014buy}
for a formulation with adverse selection.  These models describe a dealer's inventory-control problem, with
executions represented through reduced-form intensities. Whether their
quoting rules remain effective in an order-driven limit order book (LOB)
with explicit queues is not obvious a priori, and assessing this
is one objective of the paper.  These
assumptions are analytically convenient but usually go unverified on
microstructure timescales~\citep{lehalle2018market}, where flow is
episodic, persistent, and regime-dependent, and where execution
depends not only on the quote distance from the mid-price, but also on
tick discreteness, finite queue depth, and the priority rules by which
resting orders are matched~\citep{smith2003,farmer2005predictive}.
GLFT is therefore used in this paper as a strong analytical benchmark
rather than as a fully realistic structural model of market making in
such a book.

A LOB is a centralized record of all
outstanding buy and sell quotes for a given asset at an exchange,
organized by price and timestamp.  Orders sitting in the book are
called limit orders (LOs); orders that execute immediately against
existing quotes are market orders (MOs).  Most modern exchanges match
incoming flow according to price--time priority: at each price level,
the oldest resting order is executed first-in-first-out (FIFO).  A
market maker is an agent that continuously posts LOs on both
sides of the book, aiming to capture the bid--ask spread while keeping
the inventory---the net long or short position accumulated from
asymmetric fills---within tolerable bounds.  It is this LOB
market-making problem---and its robustness to the regime-switching order flow that the stylized AS/GLFT assumptions omit---that the present
paper takes as its focus.

\paragraph{Related work.}
The use of deep reinforcement learning (RL) for market making has a
growing literature that complements the closed-form controllers above
by learning quoting policies directly from interaction.
\citet{spooner2018mm} train market makers using temporal-difference
learning and the linear function approximation based on tile coding in a
historical-replay environment with simulated order execution.
\citet{kolm2020modern} explore the broader role of RL
in finance and discuss market making as a canonical application, and
\citet{zhang2020deep} apply deep RL to systematic trading in a more
general way.  \citet{gueant2019drl} apply deep RL to corporate-bond market
making, emphasizing scalability to high-dimensional inventory states.
More recently, \citet{gasperov2021signals,gasperov2021approaches}
study deep RL approaches for market making, and
\citet{sadighian2019deep} applies it to cryptocurrency market making,
while \citet{spooner2020robust} consider adversarial RL and worst-case
robustness, and \citet{ganesh2019rl} study RL market making in a
multi-agent dealer market.  \citet{barucci2026market} study a market where
informed traders react to temporary mispricing, showing that the
mispricing or its filtered estimate induces asymmetric quoting beyond
inventory effects. Their stochastic-control analysis provides a related
theoretical rationale for conditioning our learned quoting policy on
beliefs about directional order-flow regimes.
Some evaluations on historical LOB data approximate order execution
using simplified fill probabilities or slippage models. Such choices
can affect policy comparisons when queue priority is not modeled
explicitly. Moreover, a fixed order-flow history limits counterfactual
tests of persistent directional pressure.
We therefore train and evaluate both the reinforcement-learning market
maker (RLMM) and the analytical benchmarks inside the same queue-aware
event-driven simulator, using paired random seeds for the RLMM
comparisons. Their actions affect the book, so realized order-flow
paths can differ across policies.

\paragraph{Contributions.}
\begin{itemize}[leftmargin=1.2em,itemsep=1pt,topsep=2pt]
\item We calibrate a GLFT policy to a queue-faithful event-driven
  LOB simulator via censored waiting-time estimation of the fill
  intensity and a volatility signature plot for the diffusion scale,
  and assess the trade-off between mean profit and loss (PnL) and inventory control
  relative to an at-best baseline that quotes at the current best bid
  and ask.
\item We train a distributional Deep Q-Network (DQN) with a
  categorical representation of the return distribution (C51) on the same simulator and
 show empirically that it outperforms GLFT at comparable levels of
  PnL dispersion under paired-seed evaluation.
\item We identify inventory saturation under persistent directional
  order flow as the dominant failure of stationarily trained
  market-making agents and the GLFT benchmark.
\item We train a separate controller on
 regime-switching market-order flow using two
  auxiliary signals: a Bayesian online change-point filter for the
  latent directional order flow bias and a queue-adjusted
  quote-exposure imbalance for the agent's own resting quotes.  This
  pipeline substantially restores profitability under
  regime-switching order flow.
\item We introduce a scenario-bandit robust fine-tuning step that
  upweights low-return regime scenarios and show that it improves the
  regime-trained controller under random-persistence and
  correlated-direction stress tests.
\end{itemize}

The remainder of the paper is organized as follows.
Section~\ref{sec:problem_setup} introduces our event-driven
zero-intelligence (Santa Fe) LOB simulator, calibrates the GLFT policy
using censored waiting times and a volatility signature
plot, benchmarks it against the at-best baseline, and presents first
evidence that the closed form breaks under static asymmetry and
persistent directional flow.  Section~\ref{sec:learning_formulation}
places market making as a semi-Markov decision process (SMDP),
details the state, action and reward, and describes the Rainbow
distributional DQN (C51) backbone.
Section~\ref{sec:stationary_results} evaluates the stationary-flow
controller (Algorithm~A) against GLFT in terms of risk and return.
Section~\ref{sec:robustness} then introduces the regime-switching flow
model and exposes the inventory-saturation failure of this
stationarily trained controller, which
Section~\ref{sec:nonstat_training} addresses by training Algorithm~B
independently with a Bayesian change-point flow belief and a
queue-adjusted quote-exposure imbalance.
Section~\ref{sec:interpretability} examines how the learned value
function varies with inventory, spread, and inferred order-flow
conditions.
Section~\ref{sec:stress_tests} tests the controller under different
regime durations, randomly varying regime durations, and correlated
order-flow directions.
Section~\ref{sec:scenario_bandit_finetuning} presents Algorithm~C,
which fine-tunes the controller by sampling scenarios in the low-return regime 
more often. The final section discusses the results and concludes the
paper. The Appendices~\ref{app:bayesian_flow_filter}
and~\ref{app:glft_misspecification} derive the Bayesian online flow
filter and the finite-horizon GLFT misspecification curve, respectively.

\section{Problem Setup and Benchmarks}
\label{sec:problem_setup}

\paragraph{Market model.}
We use an event-driven, continuous-time, zero-intelligence (ZI) LOB
simulator in the Santa Fe tradition
\citep{smith2003,farmer2005predictive}.
Time is measured in simulation time units, denoted by $\mathrm{u}$.
We briefly summarize the
model here; a comprehensive treatment, including the
empirical motivation for each parameter definition and its calibration, can be found in
\citet{bouchaud2018trades}.  Prices live on a discrete,
equally spaced grid whose spacing is the tick size, the minimum
price change of the LOB: each grid position corresponds to exactly
one admissible quote price, so all price-valued quantities (best
bid, best ask, MM quotes) vary in integer multiples of the tick
size, while queue depths are integer counts of resting unit lots at
each level.  The state of the LOB at any instant is the vector of
outstanding volumes at each price level on the bid and ask sides.
All orders, incoming LOs, MOs, and cancelations, are of unit size,
so the order flow is fully described by event counts at each price level.
Three exogenous Poisson flows drive the book's evolution:
\begin{itemize}[leftmargin=1.2em,itemsep=1pt,topsep=2pt]
\item \textbf{Limit orders (LOs);} LOs arrive at the total rate
 $\lambda N_{\mathrm{grid}}$, with $N_{\mathrm{grid}}=50$.
  Each arrival is assigned to either side with equal probability
  and placed uniformly among that side's eligible passive price
  levels. An arriving LO joins
  the queue at its submitted price under price--time priority.
\item \textbf{Market orders (MOs).} MOs arrive as a Poisson process of
  total intensity $2\mu$; each MO is a buy with probability
  $p_{\mathrm{buy}}$ and a sell with probability $1-p_{\mathrm{buy}}$.
  An MO consumes the oldest resting order at the opposite best.

\item \textbf{Cancellations.}  Each resting LO is canceled
 independently at a rate $\theta_{\mathrm{cxl}}$ per unit size, so that the total
 cancelation rate at a given price level scales linearly with its
  depth.
\end{itemize}
The parameter triple $(\lambda, \mu, \theta_{\mathrm{cxl}})$ together with
$p_{\mathrm{buy}}$ fully specifies the stochastic flow.

\paragraph{Implementation features.}
Beyond the above described events, the simulator is characterized by two
features needed for a faithful evaluation of market-making policies:
\begin{itemize}[leftmargin=1.2em,itemsep=1pt,topsep=2pt]
\item \textbf{Per-order queue tracking,} so that each fill is
  attributed to a specific resting order rather than to the aggregate
  depth at a level, and the MM's queue position is maintained
  consistently across events.
\item \textbf{Grid re-centering,} which shifts the numerical price
  support when the mid drifts, keeping the book representation stable
  over long episodes without altering the economic state.
\end{itemize}

\paragraph{Analytical benchmarks.}
The simplest benchmark strategy is the \emph{at-best baseline}, a
naive market-making strategy that posts at and
tracks the current best bid and ask and leaves orders resting until
they are filled. The at-best strategy serves as a simple reference
baseline for both analytical and learned policies.

As the main benchmark, we implement the GLFT
policy~\citep{gueant2013dealing}, which 
 assumes that the fill intensity of a quote posted
at distance \(\delta\) in ticks from the mid decays exponentially,
\begin{equation}\label{eq:Lambda}
  \Lambda(\delta)=A e^{-\kappa \delta},
\end{equation}
where \(A>0\) is the baseline fill intensity and \(\kappa>0\) controls
how quickly the execution probability vanishes as the quote moves away
from the mid; here \(\delta\) is the quote offset relative to the
 mid-price reference, measured on the LOB price grid.  Under a constant
absolute risk aversion (CARA) objective with risk aversion
\(\gamma_{\mathrm{GLFT}} > 0\) and a Brownian mid-price with volatility
\(\sigma\), we use the GLFT asymptotic approximation to the optimal
quotes~\citep{gueant2013dealing}:
\begin{align}
  p_t^{a}
  &= m_t
  + \frac{1}{\gamma_{\mathrm{GLFT}}}\log\!\left(1 + \frac{\gamma_{\mathrm{GLFT}}}{\kappa}\right)
  - \frac{2 q_t - 1}{2}
    \sqrt{\frac{\sigma^2\gamma_{\mathrm{GLFT}}}{2\kappa A}
      \left(1 + \frac{\gamma_{\mathrm{GLFT}}}{\kappa}\right)^{\!1+\frac{\kappa}{\gamma_{\mathrm{GLFT}}}}},
  \label{eq:glft_ask}\\[4pt]
  p_t^{b}
  &= m_t
  - \frac{1}{\gamma_{\mathrm{GLFT}}}\log\!\left(1 + \frac{\gamma_{\mathrm{GLFT}}}{\kappa}\right)
  - \frac{2 q_t + 1}{2}
    \sqrt{\frac{\sigma^2\gamma_{\mathrm{GLFT}}}{2\kappa A}
      \left(1 + \frac{\gamma_{\mathrm{GLFT}}}{\kappa}\right)^{\!1+\frac{\kappa}{\gamma_{\mathrm{GLFT}}}}},
  \label{eq:glft_bid}
\end{align}
where \(m_t\) is the reference LOB mid-price at time \(t\) and
\(q_t\in\mathbb{Z}\) is the agent's signed inventory.  The GLFT
derivation treats prices as continuous; in our implementation we
discretize $p_t^{b}$ and $p_t^{a}$ by rounding the bid down and the
ask up to admissible tick indices before submitting them to the LOB.

The GLFT quotes combine spread and inventory adjustments.
The term $\tfrac{1}{\gamma_{\mathrm{GLFT}}}\log(1+\gamma_{\mathrm{GLFT}}/\kappa)$
sets the baseline half-spread (common to both sides and independent
of inventory), while the square-root term, scaled by the
inventory-dependent factor $(2q_t \mp 1)/2$, generates the
asymmetric skew that pulls both quotes toward the side on which the
MM wants to reduce its position.  The execution parameters
\(A\) and \(\kappa\) are calibrated from simulated data (see the next paragraph), and \(\gamma_{\mathrm{GLFT}}\) is the only
free control parameter that generates an explicit risk--return sweep: the larger
$\gamma_{\mathrm{GLFT}}$, the more aggressive the inventory skew, and hence the tighter
the inventory control.
GLFT thus yields interpretable, inventory-aware quotes, which makes it
a strong benchmark.  Because it does not endogenize queue position,
price--time priority, or the discrete execution mechanics of a FIFO
LOB, it serves as a disciplined analytical baseline against which the
RL controller is evaluated inside a more realistic microstructure.

\paragraph{Data and calibration.}
GLFT calibration in a FIFO LOB is itself non-trivial.  We follow the
censored waiting-time approach described by
\citet{fernandeztapia2015modeling}, adapted to our
event-driven simulator.  We use a fixed Santa Fe baseline with parameters
\begin{equation}
  \lambda = 0.06\,\mathrm{u}^{-1}, \qquad
  \mu = 0.10\,\mathrm{u}^{-1}, \qquad
  \theta_{\mathrm{cxl}} = 0.02\,\mathrm{u}^{-1},
\end{equation}
GLFT execution intensities
and volatility are calibrated from data generated by this
configuration, using episodes separate from evaluation.

\begin{figure}[htbp]
  \centering
  \includegraphics[width=0.7\linewidth]{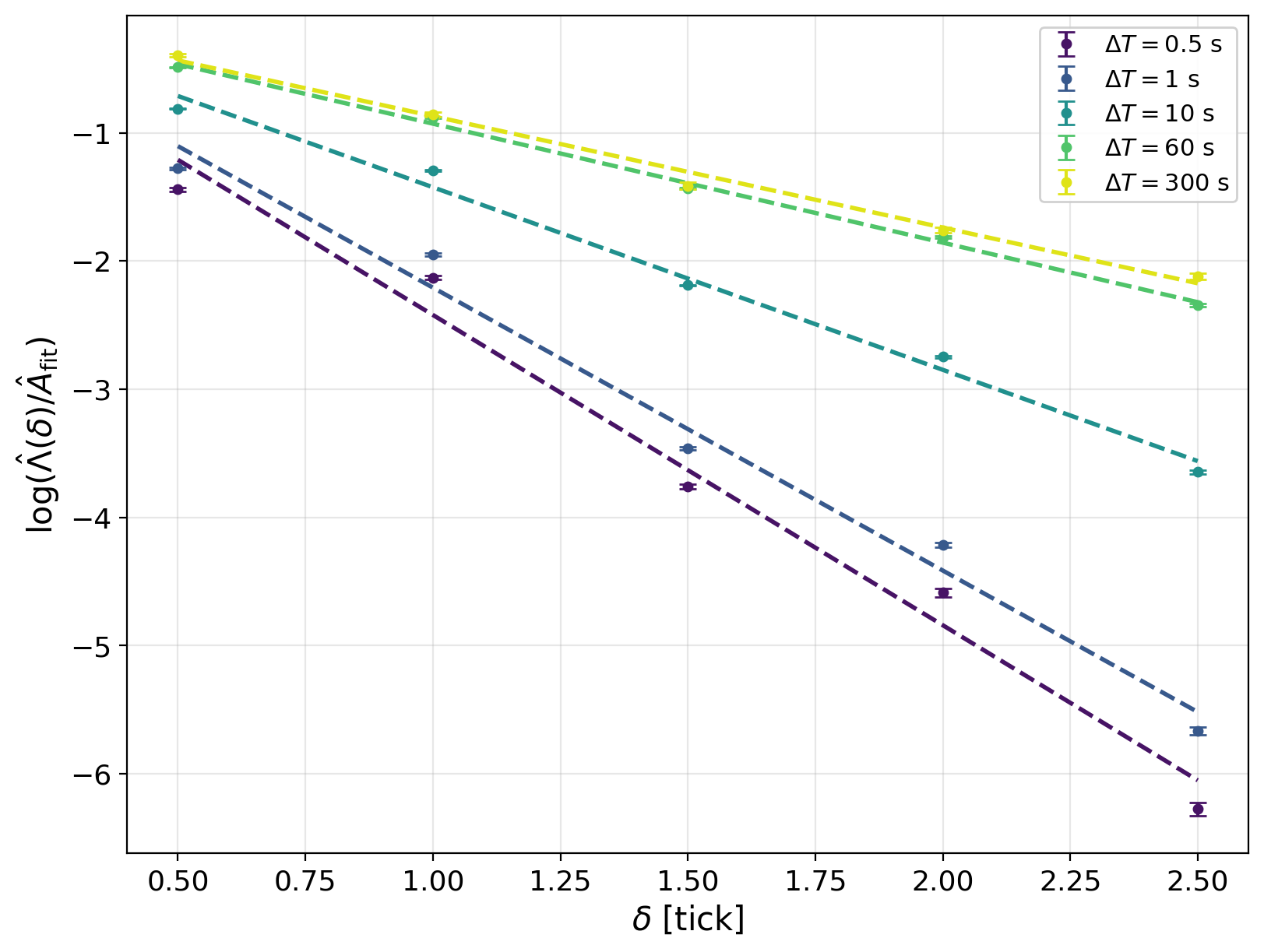}
  \caption{Censored waiting-time calibration of $\Lambda(\delta)=Ae^{-\kappa\delta}$
for five observation windows $\Delta T_{\mathrm{calib}}$.
Mean normalized log-intensities over $N_{\mathrm{rep}}=20$ independent
runs ($10^6$ LOB events per distance and run); error bars show the
standard error of the mean (SEM).
Dashed lines show log-linear fits over $0.5$--$2.5$ ticks.
Longer windows yield flatter curves, consistent with mid-price
and queue evolution weakening the dependence on initial quote distance.}
  \label{fig:glft_calib}
\end{figure}

Calibration uses stationary symmetric flow, with equal buy and sell MO
probabilities. Quote distance is measured from the submitted tick-rounded
price to the mid-price immediately before insertion.
The main input of the GLFT strategy is the fill intensity of
Eq.~\ref{eq:Lambda}. To estimate it, following the approach in
\citet{fernandeztapia2015modeling}, we proceed as follows: for each
quote distance $\delta$, a probe limit order is observed over a fixed
window of length $\Delta T_{\mathrm{calib}}$, the same for every probe.
For each probe $i$, let $\tau_i$ denote the time from posting to its
first fill. If a probe remains unfilled at the deadline, its waiting
time is right-censored, and the order is canceled and replaced at the
next eligible event. We record
\begin{equation}
    \widetilde{\tau}_i = \min(\tau_i,\Delta T_{\mathrm{calib}}),
    \qquad
    I_i = \mathbf{1}\{\tau_i < \Delta T_{\mathrm{calib}}\},
\end{equation}
where $\widetilde{\tau}_i$ is the observed exposure time and $I_i$
indicates a fill before censoring. If the simulation ends earlier,
the remaining simulation time replaces $\Delta T_{\mathrm{calib}}$
for that probe. Under the exponential-rate assumption, the
maximum-likelihood estimator of the intensity is
\begin{equation}
    \hat{\Lambda}(\delta)
    = \frac{\sum_i I_i}{\sum_i \widetilde{\tau}_i},
\end{equation}
the ratio of observed fills to total exposure time. The fitted
intensity curve depends on the observation window: longer windows
produce flatter estimates (Figure~\ref{fig:glft_calib}). We use
$\Delta T_{\mathrm{calib}}=0.5\,\mathrm{u}$ to limit changes in quote
distance during observation.

The probing scheme yields empirical intensities
$\{\hat{\Lambda}(\delta_k)\}_{k=1}^{K_\delta}$ over a finite grid
of quote distances. Pooling fills and exposure times by distance,
we fit $\hat{\Lambda}(\delta)=A_{\mathrm{fit}}e^{-\kappa\delta}$
by ordinary least squares in log-space over $0.5$--$5$ ticks,
using $\Delta T_{\mathrm{calib}}=0.5\,\mathrm{u}$.
To correct approximately for the effect of Brownian mid-price
fluctuations during the observation window, we set
$A=A_{\mathrm{fit}}x/(e^x-1)$, where
$x=\kappa^2\sigma^2\Delta T_{\mathrm{calib}}/2$. The resulting parameters are
\begin{equation}
  \hat{A} = 0.1132\ \mathrm{u}^{-1},
  \qquad
  \hat{\kappa} = 2.240\ \mathrm{ticks}^{-1},
  \label{eq:glft_calib_values}
\end{equation}
and these are the values used for the GLFT benchmark throughout the
paper, as well as in the analytical misspecification analysis of
Appendix~\ref{app:glft_misspecification}.

We estimate the volatility $\sigma$ entering the GLFT solution
from the mid-price volatility signature plot, which captures
the timescale dependence of price diffusivity.
For a mid-price series $\{m_t\}$ and a lag $h$ measured in $\mathrm{u}$,
the signature curve is
\begin{equation}
    \hat{\sigma}(h)
    = \sqrt{\frac{\mathbb{E}[(m_{t+h}-m_t)^2]}{h}},
\end{equation}
estimated empirically using overlapping increments.
For a Brownian diffusion $dS_t=\sigma\,dW_t$, the curve is
flat, whereas microstructure effects distort it at short
lags. We therefore take $\sigma$ from the long-horizon
plateau, which provides an effective diffusion scale
less affected by short-term microstructure effects.
For the baseline Santa Fe simulator with
$(\lambda,\mu,\theta_{\mathrm{cxl}})=(0.06,0.10,0.02)\,\mathrm{u}^{-1}$,
we obtain $\sigma\approx 0.30$ ticks/$\sqrt{\mathrm{u}}$,
which we use in the GLFT coefficients.

\begin{figure}[htbp]
  \centering
  \makebox[\linewidth][c]{%
  \begin{minipage}{1.18\linewidth}
  \centering
  \begin{subfigure}[t]{0.49\linewidth}
    \centering
    \includegraphics[width=\linewidth,height=6cm,keepaspectratio]{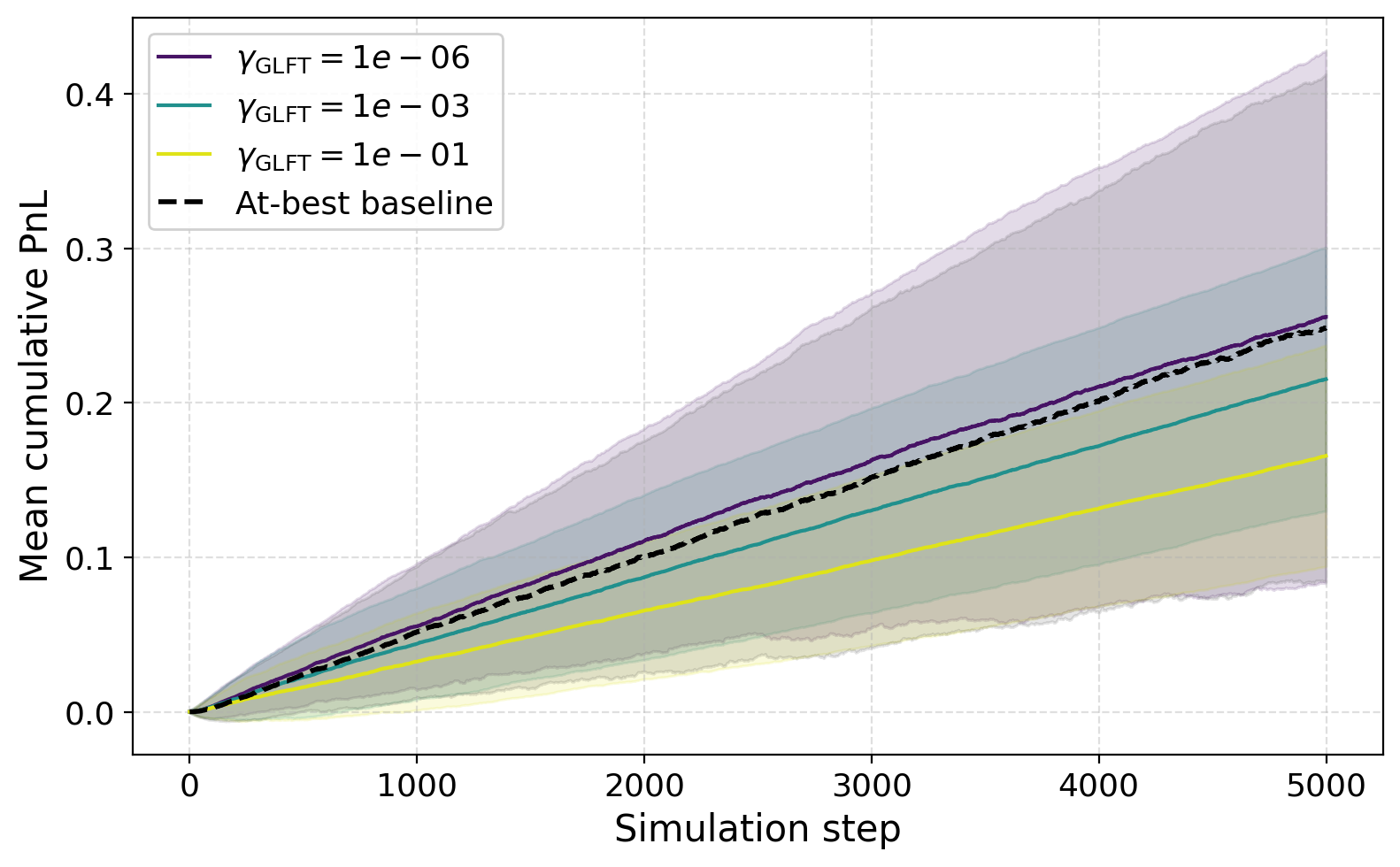}
    \caption{Mean cumulative PnL across simulations.}
    \label{fig:glft_pnl_cum}
  \end{subfigure}
  \hfill
  \begin{subfigure}[t]{0.49\linewidth}
    \centering
    \includegraphics[width=\linewidth,height=6cm,keepaspectratio]{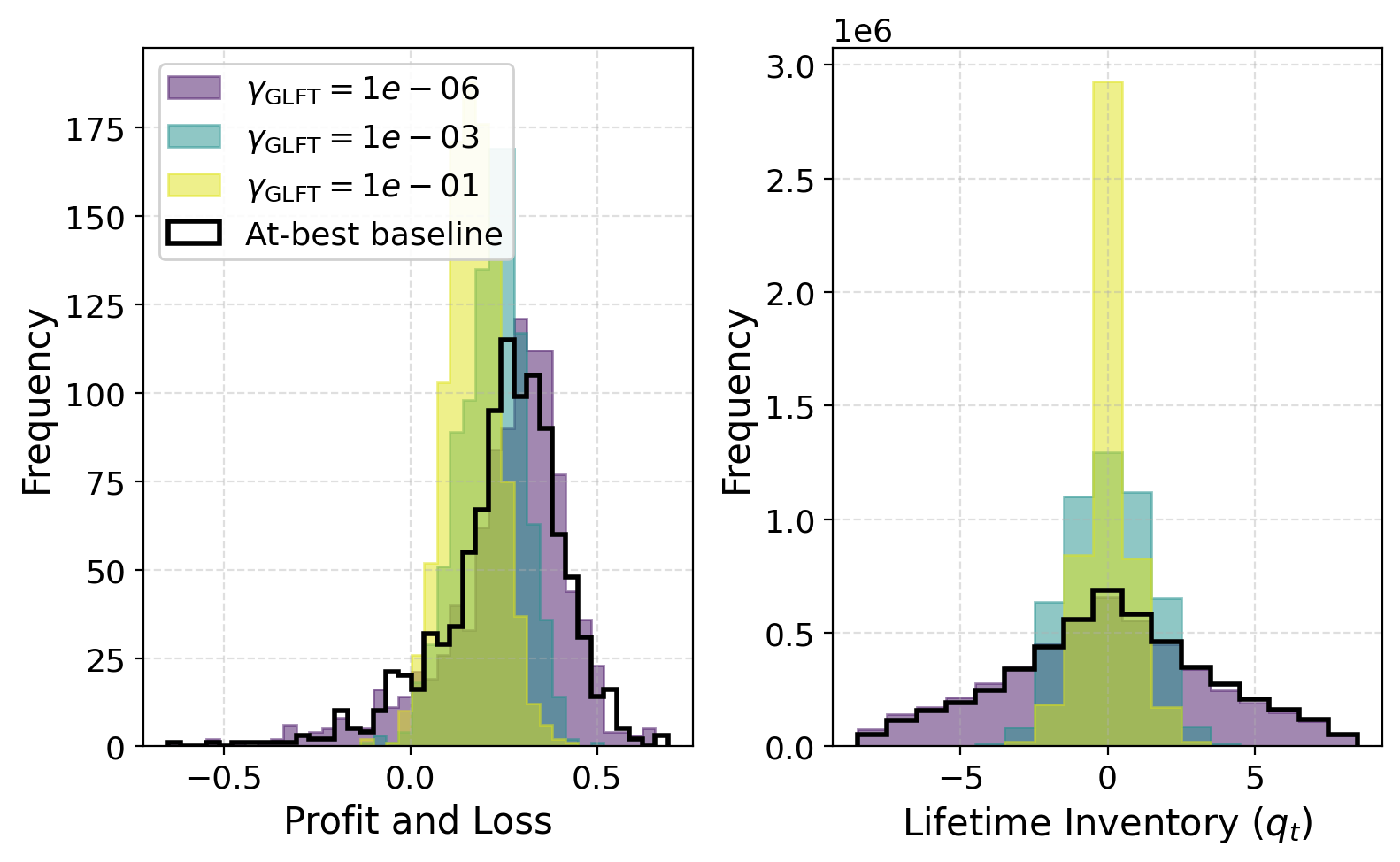}
    \caption{Terminal PnL and lifetime-inventory histograms.}
    \label{fig:glft_hist_joint}
  \end{subfigure}
  \end{minipage}}
  \caption{
    GLFT versus the at-best baseline over $N_{\mathrm{sim}}=1000$
    episodes of $N_{\mathrm{steps}}=5000$ LOB events per configuration,
    using separate seed sets and
    $\gamma_{\mathrm{GLFT}}\in\{10^{-6},10^{-3},10^{-1}\}$.
    }
  \label{fig:glft_vs_naive}
\end{figure}

\paragraph{Performance metric and analytical benchmarks.}
Throughout the paper, performance is evaluated using gross
marked-to-market PnL,
\[
  \mathrm{PnL}_T=X_T+q_Tm_T-(X_0+q_0m_0),
\]
where $X_t$ is the MM's cash balance, $q_t$ its signed inventory,
and $m_t$ the LOB mid-price. Cash and prices are expressed in
monetary units, with a tick value of $0.01$. Reported PnL excludes
fees, rebates, and terminal liquidation costs.
Figure~\ref{fig:glft_vs_naive} compares GLFT with the at-best baseline
over $1000$ episodes of $5000$ LOB events per configuration, using
separate seed sets. At $\gamma_{\mathrm{GLFT}}=10^{-6}$, GLFT has a
slightly higher sample mean PnL than the baseline. Across the three
settings, higher risk aversion lowers mean PnL and concentrates
inventory more tightly around zero.
In the continuous GLFT formula, the base half-spread tends to $1/\kappa$
and the inventory adjustment vanishes as $\gamma_{\mathrm{GLFT}}\to0$.
In the simulator, tick rounding can make different continuous quotes
produce the same submitted prices. The observed comparison therefore
reflects both the analytical inventory adjustment and its implementation
on the tick grid.

\paragraph{Where stationary quoting fails: directional order flow.}
The calibrated GLFT benchmark assumes stationary, symmetric fill
intensities, represented in our simulator by a constant buy-MO
probability $p_{\mathrm{buy}}=0.50$. In real markets, however,
order flow exhibits persistent directional imbalances, partly
associated with metaorder splitting and correlated trading
\citep{toth2015persistent,bouchaud2009digest,lillo2004longmemory}.
Recent microstructure studies also identify shifts between
intraday directional-flow regimes~\citep{tsaknaki2025online},
which are absent from the stationary benchmark.

\begin{figure}[htbp]
  \centering
  \includegraphics[width=0.7\linewidth]{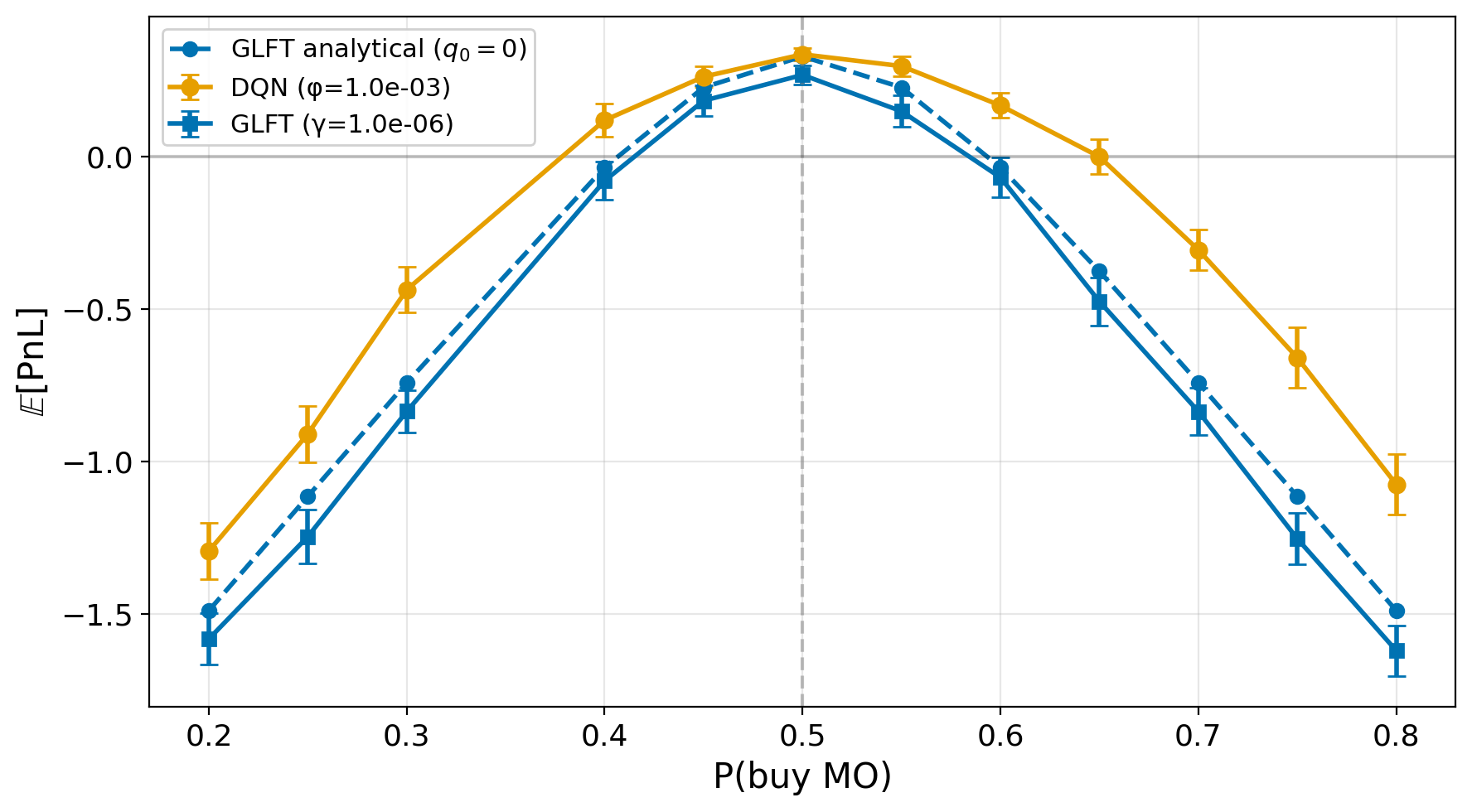}
    \caption{
    Mean terminal PnL under fixed order-flow asymmetry.
    Solid curves show RLMM (orange) and GLFT (blue) in the FIFO
    simulator over 100 episodes of 5,000 LOB events per
    $p_{\mathrm{buy}}$. Error bars indicate pointwise 95\%
    confidence intervals for the empirical means.
    The dashed curve shows the finite-horizon analytical
    approximation of Appendix~B, initialized at $q_0=0$
    with a fixed horizon $T\approx844\,\mathrm{u}$.
    }
  \label{fig:misspec_comparison}
\end{figure}

We first isolate the effect of directional asymmetry by holding
$p_{\mathrm{buy}}$ constant throughout each episode and varying
it around $0.50$. Figure~\ref{fig:misspec_comparison} shows
the resulting deterioration in GLFT performance as a function
of $p_{\mathrm{buy}}$. Mean terminal PnL peaks at
$p_{\mathrm{buy}}=0.50$ and becomes negative once
$|p_{\mathrm{buy}}-0.50|\gtrsim 0.08$.
We also report a finite-horizon analytical approximation,
initialized at zero inventory and derived in Appendix~B.

Figure~\ref{fig:misspec_comparison} also shows that the RLMM
introduced in Sections~3--4 remains profitable on average
over a wider range of flow asymmetries
($p_{\mathrm{buy}}\in[0.40,0.60]$, approximately),
although its performance also deteriorates as asymmetry increases.

\begin{figure}[htbp]
  \centering
  \includegraphics[width=0.7\linewidth]{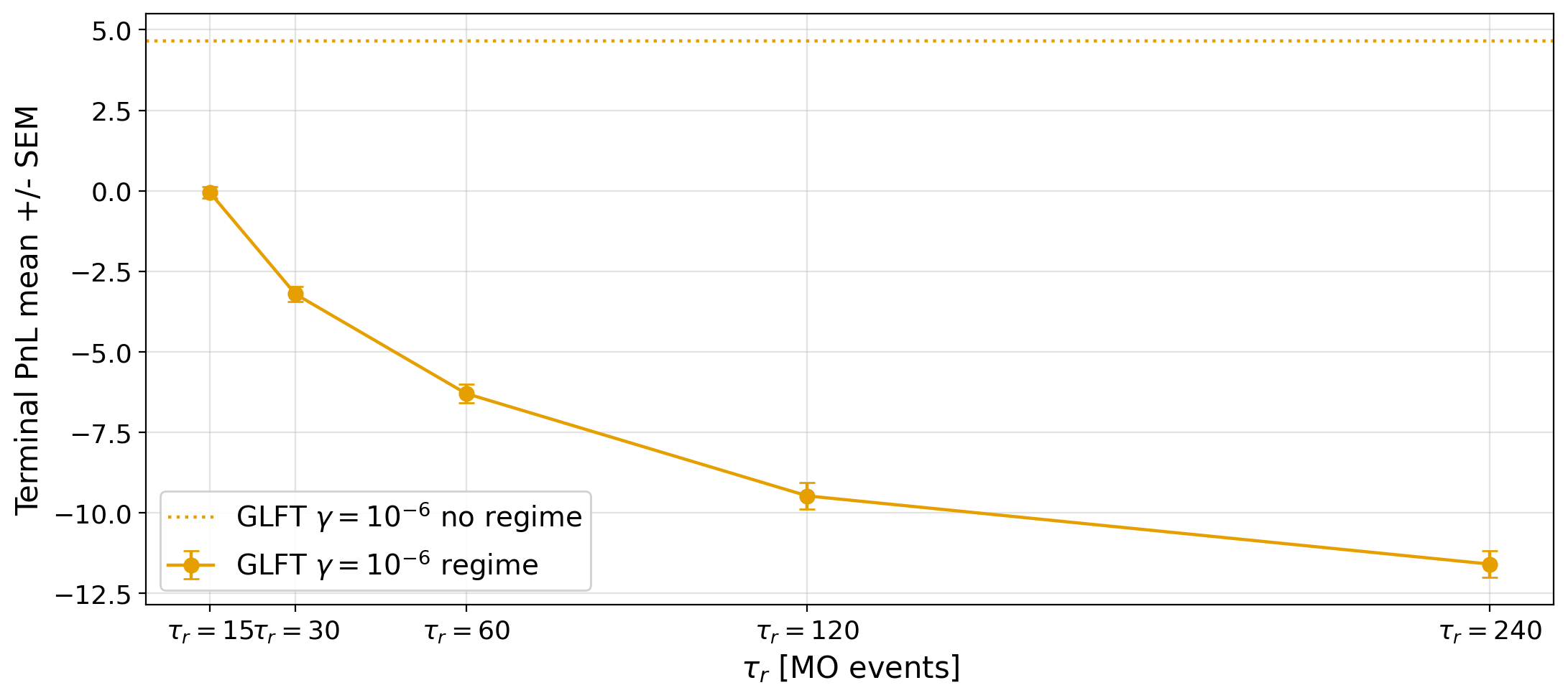}
    \caption{
    GLFT mean terminal PnL versus regime persistence $\tau_r$
    (MO clock), with standard-error bars.
    Regime durations are exponentially distributed with mean
    $\tau_r$ MO events, and
    $p_{\mathrm{buy}}\sim\mathrm{Uniform}[0.20,0.80]$
    ($\omega=0.30$; model in Section~5).
    Results use 100 episodes of 100,000 LOB events per setting.
    The dotted line marks the symmetric no-regime baseline
    ($\omega=0$).
    }
  \label{fig:glft_degradation_tau}
\end{figure}

We next examine GLFT under regime-switching order flow.
Figure~\ref{fig:glft_degradation_tau} evaluates the calibrated
policy with piecewise-constant buy-MO probabilities and
exponentially distributed regime durations of mean $\tau_r$
MO events (the model is detailed in Section~\ref{sec:robustness}).
Mean terminal PnL is near zero at $\tau_r=15$ and becomes
increasingly negative as regimes lengthen. Thus, GLFT loses
profitability under persistent directional flow, which its
quoting rule does not explicitly capture. This failure
motivates the remainder of the paper: Sections~\ref{sec:learning_formulation}
and~\ref{sec:stationary_results} establish the RL controller's
advantage over GLFT under stationary flow, while
Sections~\ref{sec:robustness}--\ref{sec:scenario_bandit_finetuning}
examine its regime fragility and address it through regime-aware
training and scenario-bandit fine-tuning.

\section{Reinforcement Learning Formulation}
\label{sec:learning_formulation}

We formulate market making as a reinforcement-learning problem
and describe the RL controllers evaluated against GLFT
(Section~\ref{sec:problem_setup}). Although the simulator
evolves event by event, the MM updates its quotes only at
decision times. At each step, the agent observes a compact
state $s_t$ (Eq.~\ref{eq:state_vector}), selects a discrete
quoting action and receives the reward accumulated until
the next decision. The state summarizes the bid--ask spread,
queue depths at and near the best quotes, and inventory,
with additional flow-related features in the regime-switching
experiments. Although the full simulator is Markovian, this
compact observation does not reveal the latent directional
regime. Section~\ref{sec:nonstat_training} addresses this by
incorporating the history-dependent flow belief $b_t$
(Eq.~\eqref{eq:phaseb_belief_summary}) and the
queue-exposure signal $x_t$.

\paragraph{Decision clock (throttle).}
The MM does not update its quotes at every LOB event. Instead,
a time throttle $\tau_c$ allows a new decision to be made on the first
event that occurs at least $\tau_c$ after
the previous decision. We index decisions by $t\in\mathbb{N}$
and fix $\tau_c=1\,\mathrm{u}$.
At the baseline rates of Section~\ref{sec:problem_setup},
the simulator generates approximately $3.0$ LO arrivals,
$0.2$ MO arrivals, and $2.8$ cancelations per $\mathrm{u}$,
consistent with the steady-state balance between order
arrivals and removals. Thus, roughly six LOB events occur
per $\mathrm{u}$ on average.

Between decisions $t$ and $t+1$, the simulator processes
a random number $N_t\geq1$ of book events, indexed by
$n=0,\ldots,N_t-1$, each generating an event-level reward
$r_{t,n}$ (Eq.~\eqref{eq:reward}).
To maintain consistent discounting on the event clock,
we aggregate these rewards as
\begin{equation}
    r_t=\sum_{n=0}^{N_t-1}\gamma^n r_{t,n},
    \label{eq:smdp_reward}
\end{equation}
where $\gamma$ is the discount factor per-event.

Because $N_t$ varies between decisions, the process at
decision times has a semi-Markov structure at the full
level of the simulator-state 
\citep{puterman1994,sutton1999options}.
The next-state value is discounted by
$\Gamma_t=\gamma^{N_t}$, resulting in the one-step target
$r_t+\Gamma_t\max_{a'}Q(s_{t+1},a')$ used by the controller.
When $N_t\equiv1$, we recover $r_t=r_{t,0}$ and
$\Gamma_t=\gamma$, as in standard one-step Q-learning.
Discounting is thus defined on the event clock, while
$\tau_c$ determines the decision frequency in the simulator
time.

\paragraph{State and action space.}
The MM observes the state vector
\begin{equation}
  s_t=
  \bigl[\mathrm{spr}_t,\;
  d_{t,0}^{b},\ldots,d_{t,K}^{b},\;
  d_{t,0}^{a},\ldots,d_{t,K}^{a},\;
  q_t,\;e_t\bigr],
  \label{eq:state_vector}
\end{equation}
where $\mathrm{spr}_t$ is the bid--ask spread;
$d_{t,k}^{b}$ and $d_{t,k}^{a}$ denote queue depths
at offset $k$ from the best bid and ask, respectively,
with $K=1$ in all experiments and zero depth at empty levels.
The variable $q_t$ is the signed inventory, while $e_t$
contains auxiliary features used only under regime-switching
flow. The simulator tracks the MM's resting orders and
their FIFO positions internally, although these are not
directly observed by the stationary controller.

In the stationary experiments of
Section~\ref{sec:stationary_results}, $e_t$ is disabled.
For the regime-switching controller of
Section~\ref{sec:nonstat_training}, it comprises two
complementary signals: a Bayesian estimate of directional
order flow and a measure of the MM's own quote exposure,
defined below.

The first signal is obtained from a Bayesian online
change-point filter~\citep{adams2007bocpd} applied to the
signed market-order stream. At each MO arrival, the filter
updates the posterior over the latent buy-MO probability
and regime run length. The controller receives the belief
summary
\begin{equation}
  b_t = \bigl(\widehat{\iota}_t,\;\bar{\ell}_t\bigr),
  \label{eq:phaseb_belief_summary}
\end{equation}
where $\widehat{\iota}_t=2\widehat p_t-1$ is the estimated
directional bias, $\widehat p_t$ is the posterior mean buy-MO
probability, and $\bar{\ell}_t$ is the posterior expected
run length. The latter is log-scaled and normalized before
entering the network (Appendix~\ref{app:bayesian_flow_filter}).
The filter updates on MO arrivals, independently of the
MM's decision clock. At each decision, the agent uses the
latest available belief, carried forward if no new MO
has arrived.

The second signal captures the MM's exposure through its resting
quotes. The agent maintains at most one unit-size order per side.
Let $\mathrm{qa}_t^b$ and $\mathrm{qa}_t^a$ denote the number of
units ahead of its bid and ask in their respective FIFO queues.
We define the side-specific quote-exposure proxies as
\begin{equation}
  O_t^{b}=
  \frac{\mathbf 1_{\{\text{bid alive}\}}}{1+\mathrm{qa}_t^b},
  \qquad
  O_t^{a}=
  \frac{\mathbf 1_{\{\text{ask alive}\}}}{1+\mathrm{qa}_t^a},
\end{equation}
with absent orders assigned zero exposure. The resulting
quote-exposure imbalance is
\begin{equation}
  x_t
  =
  \frac{O_t^{b}-O_t^{a}}{1+O_t^{b}+O_t^{a}}.
  \label{eq:quote_exposure_imbalance}
\end{equation}

With unit-size quotes, $x_t\in[-1/2,1/2]$: it equals
$+1/2$ ($-1/2$) when only a front-of-queue bid (ask)
is active, and approaches zero when exposures are
balanced or both orders are deep in their queues.
Unlike the MO-updated belief $b_t$, the exposure signal
$x_t$ is computed from the current queue state at each
MM decision. The auxiliary feature vector is therefore
$e_t=(b_t,x_t)$.

Each action specifies bid and ask tick offsets $(\Delta_b,\Delta_a)$
relative to the current best quotes:
\begin{equation}
    p_t^b = B_t-\Delta_b,
    \qquad
    p_t^a = A_t+\Delta_a,
\end{equation}
where $B_t$ and $A_t$ denote the best bid and ask at decision
time $t$. For a newly submitted order, a zero offset joins
the queue at the best price, a positive offset places it deeper
in the book, and an offset of $-1$ improves the best quote by
one tick when the spread permits. The same convention applies
to both sides. Thus, each action controls the quoted spread
$\mathrm{spr}_t+\Delta_b+\Delta_a$ and its asymmetry
$\Delta_a-\Delta_b$, allowing inventory-dependent quoting.

We restrict the action space to six offset pairs:
\begin{equation}
    \mathcal{A}=\bigl\{
    (-1,-1),\,(-1,0),\,(0,-1),\,
    (0,0),\,(0,+1),\,(+1,0)
    \bigr\}.
\end{equation}
The first three actions improve at least one best quote when
the spread allows. The set excludes simultaneous widening
$(+1,+1)$ and the two extreme skews $(+1,-1)$ and $(-1,+1)$.
With both quotes active, the nominal spread ranges from
$\mathrm{spr}_t-2$ to $\mathrm{spr}_t+1$ ticks, subject to
the non-crossing constraints described below, while
$\Delta_a-\Delta_b\in\{-1,0,+1\}$.
This compact action space reduces the number of
action-dependent return distributions learned by the network.

\paragraph{Quote management.}
At each decision time, the MM selects an action and adjusts
its quotes as needed. When the spread is one tick, a requested
one-tick improvement is replaced by a quote at the current
best price on that side. A non-crossing constraint also
keeps the ask at least one tick above the bid.
The hard inventory bound is enforced by suppressing the bid
at $q_t=q_{\max}$ and the ask at $q_t=-q_{\max}$.
At either limit, only three effective actions remain,
corresponding to the offsets $\{-1,0,+1\}$ on the
inventory-reducing side. The same restriction applies
to the Bellman target maximization. All six actions
remain available otherwise.

Resting orders persist between decisions unless filled.
At each update, an existing order retains its FIFO priority
if its price matches the target; otherwise, it is canceled
and reposted at the new price, joining the back of the queue.
This preserves queue priority whenever possible, avoiding
unnecessary cancellations and reposting.

\paragraph{Action-space ablation.}
We compare the six-action set with the full nine-action grid
under matched experimental conditions. Under symmetric flow,
mean terminal PnL is similar ($0.339$ vs.\ $0.331$ over
$N_{\mathrm{sim}}=300$ paired evaluation episodes, within
one standard error), with the three additional actions
selected only $10.3\%$ of the time.
Under regime-switching training, however, the nine-action
variant underperforms the six-action controller, achieving
a lower peak moving-average terminal PnL
($-0.005$ vs.\ $0.097$) over $4000$ training episodes.
This suggests that the reduced action space may also
facilitate learning under regime-switching flow.

\paragraph{Reward.}
We define the reward at the book-event level and aggregate
it over each decision interval. For event $n$ within
interval $t$, the reward consists of execution gains
measured relative to the post-event mid-price, minus
quadratic inventory penalties:
\begin{equation}
  \label{eq:reward}
  r_{t,n}
  =
  \sum_{f\in\mathcal{F}_{t,n}}
  \Delta q(f)\bigl(m^{+}_{t,n}-p(f)\bigr)
  -
  \varphi\bigl(q^{+}_{t,n}\bigr)^2
  -
  \varphi\bigl(|q^{+}_{t,n}|-q_w\bigr)_+^2.
\end{equation}
Here, $\mathcal{F}_{t,n}$ denotes the set of MM fills
at event $n$, containing at most one fill in our
simulator. The variables $m^+_{t,n}$ and $q^+_{t,n}$
denote the post-event mid-price and inventory,
respectively. The signed fill quantity $\Delta q(f)$
is positive for purchases and negative for sales,
and $p(f)$ is the execution price. Prices are expressed in ticks.

The decision-level reward $r_t$ is the discounted sum
of these event-level contributions, as defined in
Eq.~\eqref{eq:smdp_reward}.\footnote{We index the rewards by
the decision at which the action is taken: $r_t$ is achieved
during the transition from $s_t$ to $s_{t+1}$, corresponding
to $R_{t+1}$ in the convention of \citet{suttonbarto2018}.}

The coefficient $\varphi\geq0$ controls the trade-off between
execution gains and inventory risk. In addition to the quadratic
inventory penalty, an additional penalty applies when
$|q^+_{t,n}|>q_w$, where $(u)_+=\max(u,0)$.
We set $q_w=q_{\max}/2$, while the hard inventory limit
$|q^+_{t,n}|\leq q_{\max}$ is enforced by restricting
the placement of orders at inventory limits.

The MM is trained using temporal-difference (TD) learning
\citep{sutton1988td,watkins1992qlearning}, with discounting
defined on the LOB event clock. For a decision-level transition
$(s_t,a_t,r_t,s_{t+1},N_t)$, the effective discount is
$\Gamma_t=\gamma^{N_t}$, as defined above.
In Double DQN~\citep{vanhasselt2016}, the next action is
selected using the online network $Q_\theta$ and evaluated
using the target network $Q_{\theta^-}$. The resulting
one-step target is
\begin{equation}
  G_t^{(1)}
  =
  r_t+(1-d_t)\Gamma_t
  Q_{\theta^-}\!\left(
    s_{t+1},
    \arg\max_{a'\in\mathcal{A}(s_{t+1})}
    Q_\theta(s_{t+1},a')
  \right),
\end{equation}
where $d_t=1$ for terminal transitions and
$\mathcal{A}(s_{t+1})$ denotes the actions permitted by
the inventory constraint. The target-network parameters
$\theta^-$ are periodically synchronized with $\theta$.
The distributional extension is described below.

\paragraph{Rainbow DQN.}
We adopt a Rainbow-style distributional DQN~\citep{hessel2018}
to learn the MM's quoting policy. The controller combines
DQN~\citep{mnih2015}, Double DQN~\citep{vanhasselt2016},
dueling networks~\citep{wang2016dueling}, prioritized
experience replay (PER)~\citep{schaul2016prioritized},
NoisyNet exploration~\citep{fortunato2018noisy},
$n$-step returns, and C51~\citep{bellemare2017}.
Since fills occur intermittently, execution rewards are
sparse and realized returns can vary substantially.
Inventory penalties accrue at every book event.
C51 captures the distribution of discounted returns
rather than only their expected value.

Unlike standard DQN, C51 represents the return distribution
for each state--action pair on a fixed support
$[V_{\min},V_{\max}]$, discretized into $N$ equally spaced
values, called atoms:
\begin{equation}
  z_k=V_{\min}+(k-1)\Delta z,
  \qquad
  \Delta z=\frac{V_{\max}-V_{\min}}{N-1},
  \quad k=1,\ldots,N.
\end{equation}
A shared representation $h_t=\phi_\theta(s_t)$ feeds
value and advantage streams that produce logits for each
action and return atom:
\begin{equation}
  \ell_{\theta,k}(s_t,a)
  =
  V_{\theta,k}(h_t)
  +
  A_{\theta,k}(h_t,a)
  -
  \frac{1}{|\mathcal A|}
  \sum_{a'\in\mathcal A} A_{\theta,k}(h_t,a').
\end{equation}
Here, $\theta$ denotes the network parameters and
$|\mathcal A|$ the number of discrete actions.
A softmax over the atoms yields probabilities
$\pi_{\theta,k}(s,a)$, defining the expected action value
\begin{equation}
  Q_\theta(s,a)
  =
  \sum_{k=1}^{N} z_k\,\pi_{\theta,k}(s,a).
\end{equation}
Thus, actions are selected using expected returns, while
the network learns their categorical distributions.

To propagate reward information across multiple decisions,
we use $n$-step returns~\citep{suttonbarto2018}.
Since each decision interval spans a variable number
$N_t$ of book events, the cumulative event count from
decision $t$ to decision $t+i$ is
\begin{equation}
  M_i=\sum_{j=0}^{i-1}N_{t+j},
  \qquad M_0=0.
\end{equation}
The corresponding event-discounted Double-DQN target
generalizes the one-step target defined above:
\begin{equation}
  G_t^{(n)}
  =
  \sum_{i=0}^{n-1}\gamma^{M_i}r_{t+i}
  +
  (1-d_t^{(n)})\gamma^{M_n}
  Q_{\theta^-}\!\left(
    s_{t+n},
    \arg\max_{a'\in\mathcal A(s_{t+n})}
    Q_\theta(s_{t+n},a')
  \right),
\end{equation}
where $\mathcal A(s_{t+n})$ contains the actions permitted
by the inventory constraint. For terminal samples,
$n$ is the number of available transitions and
$d_t^{(n)}=1$, removing the bootstrap contribution.
When $N_{t+j}\equiv1$, the expression reduces to the
standard $n$-step target; otherwise, $\gamma^{M_i}$
accounts for the varying number of events between decisions.

The C51 target distribution is constructed by applying
the $n$-step Bellman update to the atoms of the next-state
distribution, with the next action selected by the online
network and evaluated by the target network.
The shifted probability masses are projected back onto
the fixed support $\{z_1,\ldots,z_N\}$.
In our implementation, the accumulated discounted reward
is additionally clipped to $[V_{\min},V_{\max}]$
before adding the bootstrap contribution.
Denoting the projected target probabilities by
$\pi_k^{\mathrm{target}}$, the network minimizes the
categorical cross-entropy loss
\begin{equation}
  \mathcal L_{\mathrm{C51}}
  =
  -\sum_{k=1}^{N}
  \pi_k^{\mathrm{target}}
  \log\pi_{\theta,k}(s_t,a_t).
\end{equation}

Prioritized experience replay assigns higher sampling
priority to transitions with larger prediction errors,
measured by the Kullback--Leibler (KL) divergence
between the target and predicted distributions:
\begin{equation}
  D_{\mathrm{KL}}(\pi^{\mathrm{target}}\|\pi_\theta)
  =
  \sum_{k=1}^{N}
  \pi_k^{\mathrm{target}}
  \log\left(
    \frac{\pi_k^{\mathrm{target}}}
         {\pi_{\theta,k}(s_t,a_t)}
  \right).
\end{equation}
Each replay priority $p_i$ is updated from this
divergence, subject to an adaptive cap that limits
the influence of extreme values.
\footnote{Replay priorities are capped at three times an
exponentially smoothed mini-batch median of KL divergences,
using a smoothing coefficient of $0.01$.}
Transitions are sampled according to
\begin{equation}
  P(i)=
  \frac{(p_i+\varepsilon_0)^\alpha}
       {\sum_j(p_j+\varepsilon_0)^\alpha},
  \qquad
  w_i\propto\bigl(|\mathcal B|P(i)\bigr)^{-\beta},
\end{equation}
where $\varepsilon_0>0$ ensures nonzero sampling
probabilities, $\alpha\in[0,1]$ controls prioritization,
and $|\mathcal B|$ is the replay-buffer size.
The importance-sampling weights $w_i$ correct for
nonuniform sampling in the training loss, with
$\beta\in[0,1]$ controlling the correction strength.

Exploration uses NoisyLinear layers, which introduce
learned parameter noise:
\begin{equation}
  y=
  (\mu^W+\sigma^W\odot\varepsilon^W)x
  +
  (\mu^b+\sigma^b\odot\varepsilon^b),
\end{equation}
where $\mu^W,\mu^b$ are deterministic weights and biases,
$\sigma^W,\sigma^b$ are learned noise scales,
$\varepsilon^W,\varepsilon^b$ are random perturbations,
and $\odot$ denotes element-wise multiplication.

The main architectural and optimization hyper-parameters
are summarized in Table~\ref{tab:dqn_hyperparams}.

\begin{table}[!htbp]
  \centering
  \caption{Main hyperparameters of the Rainbow-style DQN
controllers used in the reported experiments.}
  \label{tab:dqn_hyperparams}
  \small
  \begin{tabular}{@{}lccc@{}}
    \toprule
    Component & Algorithm~A & Algorithm~B & Algorithm~C \\
    \midrule
    Feature trunk & \multicolumn{3}{c}{One hidden layer with 256 units} \\
    Activation & \multicolumn{3}{c}{ReLU} \\
    Value/action head & \multicolumn{3}{c}{Dueling C51; six quote-offset actions} \\
    Noisy layers during training & Heads only & \multicolumn{2}{c}{Hidden layer + heads} \\
    C51 atoms & 51 & 101 & 101 \\
    C51 support & $[-2,2]$ & $[-3,3]$ & $[-3,3]$ \\
    Discount per event, $\gamma$ & $0.97$ & $0.999$ & $0.999$ \\
    TD horizon (decisions), $n$ & 5 & 3 & 3 \\
    Optimizer & \multicolumn{3}{c}{AdamW} \\
    Initial learning rate & $1.5\times10^{-3}$ & $1.5\times10^{-3}$ & $3\times10^{-4}$ \\
    Batch size & 128 & 64 & 64 \\
    Prioritized replay capacity & $1.5\times10^5$ & $10^5$ & $10^5$ \\
    Target-network update (gradient steps) & \multicolumn{3}{c}{2000} \\
    Training episodes & 2000 & 4000 & 2000 (fine-tuning from B) \\
    \bottomrule
  \end{tabular}
\end{table}
\FloatBarrier

\section{Stationary Order Flow Results}
\label{sec:stationary_results}

We compare the stationarily trained RLMM (Algorithm~A)
with the calibrated GLFT policy and the at-best baseline
under stationary, symmetric order flow.
Policies are evaluated using paired random seeds;
however, their interactions with the LOB may produce
different realized order-flow paths.

\begin{figure}[htbp]
  \centering
  \begin{subfigure}[b]{0.49\linewidth}
    \centering
    \includegraphics[width=\linewidth]{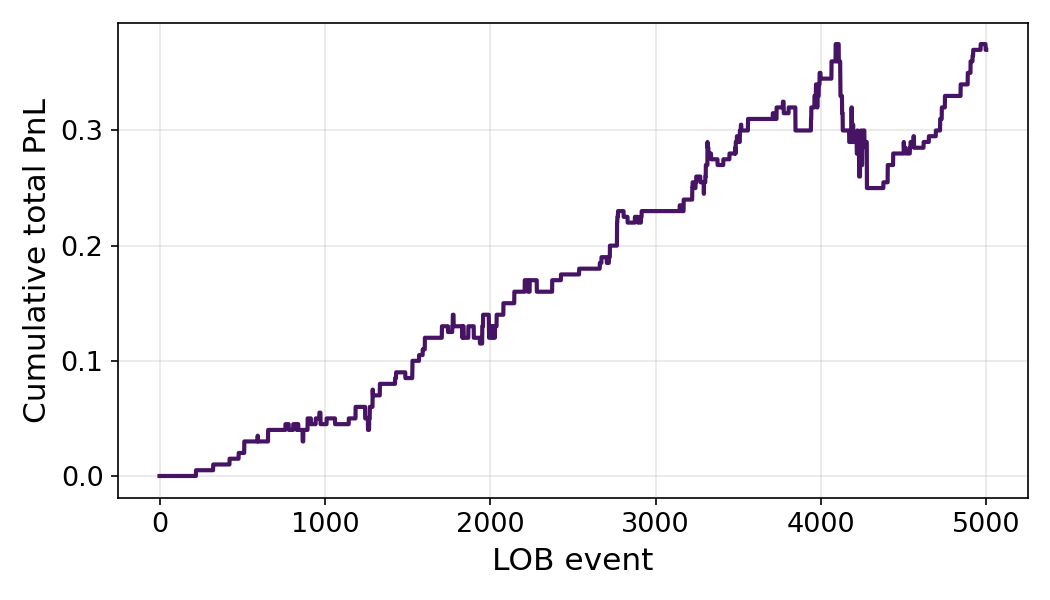}
    \caption{Cumulative total PnL.}
    \label{fig:dqn_pnl}
  \end{subfigure}
  \hfill
  \begin{subfigure}[b]{0.49\linewidth}
    \centering
    \includegraphics[width=\linewidth]{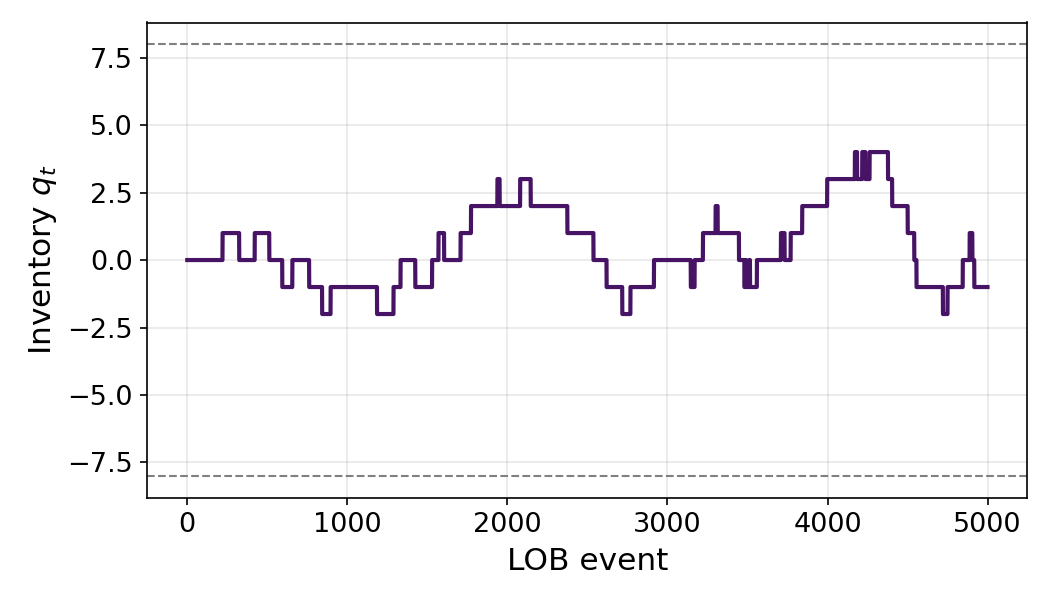}
    \caption{Inventory trajectory.}
    \label{fig:dqn_inv}
  \end{subfigure}
  \caption{Example episode under symmetric order flow, with inventory
  penalty $\varphi=10^{-3}$.
  (\subref{fig:dqn_pnl}) Cumulative PnL.
  (\subref{fig:dqn_inv}) Inventory trajectory.}
  \label{fig:dqn_episode}
\end{figure}

Figure~\ref{fig:dqn_episode} shows one evaluation episode of the
trained controller under symmetric order flow.

\begin{figure}[htbp]
  \centering
  \includegraphics[width=0.7\linewidth]{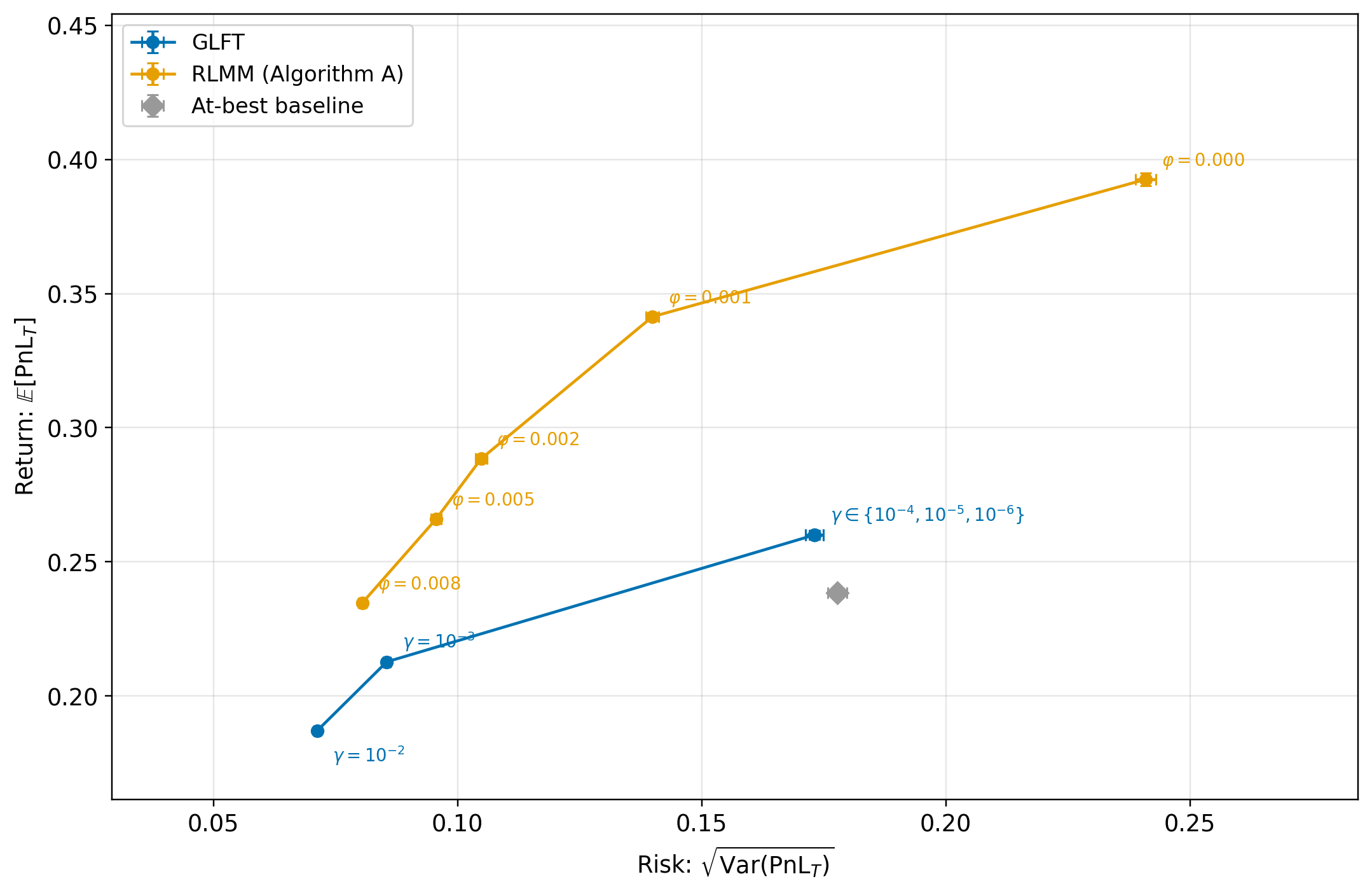}
    \caption{
    Risk--return frontier over $N_{\mathrm{sim}}=10\,000$
    paired-seed episodes of $5\,000$ LOB events, following
    $1\,000$ warm-up events, with evaluation seeds excluded
    from training. GLFT (blue) sweeps $\gamma_{\mathrm{GLFT}}$,
    RLMM (orange) sweeps $\varphi$, and the at-best baseline
    (gray) serves as reference. Markers indicate parameter
    values. Error bars show bootstrap standard errors of
    mean PnL (vertical) and PnL standard deviation
    (horizontal), based on $10\,000$ paired resamples.
    }
  \label{fig:dqn_frontier}
\end{figure}

Since the GLFT risk-aversion parameter $\gamma_{\mathrm{GLFT}}$
and the RLMM inventory penalty $\varphi$ are not directly
comparable, we evaluate both policies across their respective
parameter ranges and compare mean terminal PnL against its
standard deviation. Figure~\ref{fig:dqn_frontier}
reports results from $N_{\mathrm{sim}}=10\,000$ paired-seed
evaluation episodes per configuration, using seeds excluded
from training and fixed policy checkpoints.
Sweeping $\gamma_{\mathrm{GLFT}}$ for GLFT (blue) and
$\varphi$ for RLMM (orange), with the at-best baseline
(gray) as reference, we find that RLMM achieves higher
mean PnL at comparable levels of dispersion across the
overlapping risk range.

Unless otherwise stated, Algorithm~A refers to the
stationary-flow controller trained with $\varphi=10^{-3}$.

\section{The Robustness Problem}
\label{sec:robustness}

\paragraph{Motivation: from static asymmetry to regimes.}
Section~\ref{sec:problem_setup} showed that both GLFT and the
stationarily trained RLMM deteriorate under fixed order-flow
asymmetry (Figure~\ref{fig:misspec_comparison}), while GLFT
performance declines as directional regimes become more
persistent (Figure~\ref{fig:glft_degradation_tau}).
These results motivate testing whether the stationary RLMM's
advantage over GLFT persists under regime-switching flow.
We now specify the regime model, examine the failure mechanism
of Algorithm~A, and motivate the regime-aware training
introduced in Section~\ref{sec:nonstat_training}.

\paragraph{Regime generation.}
We model directional order flow through a piecewise-constant
buy-MO probability $p_{\mathrm{buy}}$, with regime changes
occurring on the MO clock. Regime durations are measured
in market orders, with their duration in simulator time determined by
the MO arrival rate $2\mu$. For each regime $k$, we draw
\begin{equation}
  L_k=\lceil X_k\rceil,\qquad
  X_k\sim\mathrm{Exp}(1/\tau_r),
\end{equation}
where $\tau_r>0$ approximately determines the mean regime
duration in MO events. The buy-MO probability is sampled
independently for each regime:
\begin{equation}
  p_{\mathrm{buy},k}
  \sim\mathrm{Uniform}[0.50-\omega,\;0.50+\omega].
\end{equation}
The parameter $\omega\in[0,\omega_{\max}]$ controls the
magnitude of directional imbalance. At $\omega=0$, the
model reduces to stationary symmetric flow. Throughout
the regime-switching experiments, we fix $\omega=0.30$,
giving $p_{\mathrm{buy}}\in[0.20,0.80]$, and vary $\tau_r$
to control regime persistence. Larger $\tau_r$ produces
longer-lived directional pressure.

\begin{figure}[htbp]
  \centering
  \includegraphics[width=0.85\linewidth]{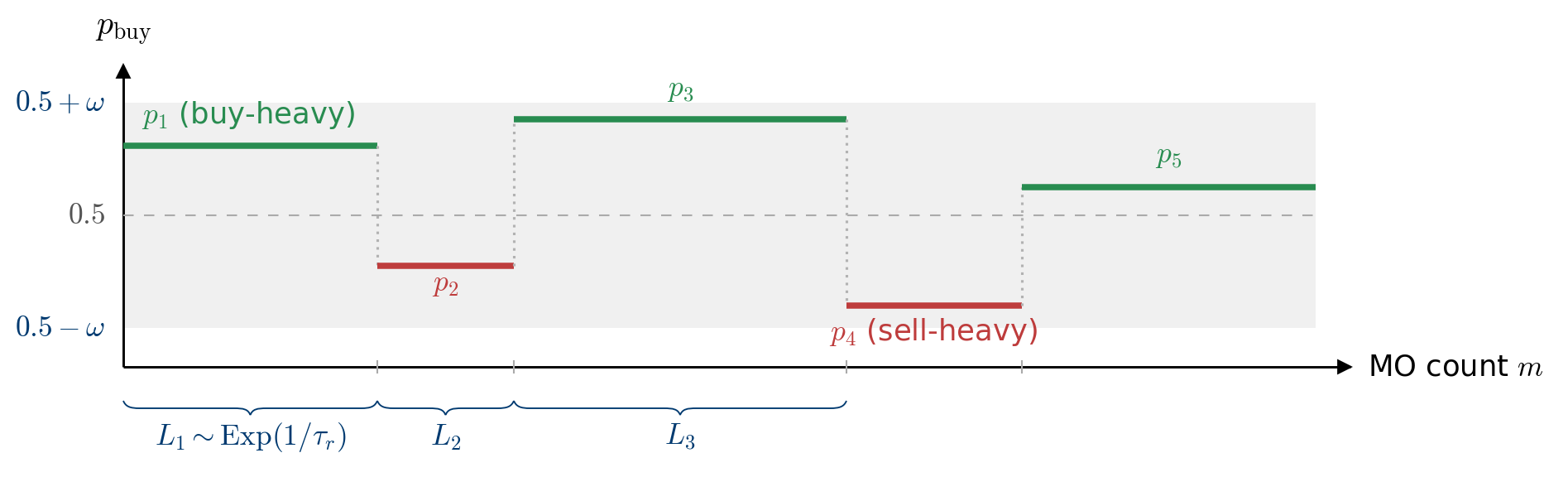}
    \caption{
    Regime-switching order flow on the MO clock.
    Within each regime $k$, the buy-MO probability remains constant,
    with $p_{\mathrm{buy},k}\sim
    \mathrm{Uniform}[0.50-\omega,0.50+\omega]$.
    Regime durations follow $L_k=\lceil X_k\rceil$,
    where $X_k\sim\mathrm{Exp}(1/\tau_r)$.
    Green (red) indicates buy-heavy (sell-heavy) regimes,
    and the shaded band spans $2\omega$.
    Throughout the experiments, $\omega=0.30$.
    }
  \label{fig:regime_model}
\end{figure}

The regime schedule $p_{\mathrm{buy}}(t)$ is hidden from
the agent, which estimates directional flow through its
Bayesian belief $b_t$ and monitors its own quote exposure
through $x_t$. Figure~\ref{fig:regime_model} illustrates a realization of this regime-switching process.

\begin{figure}[htbp]
  \centering
  \includegraphics[width=0.8\linewidth]{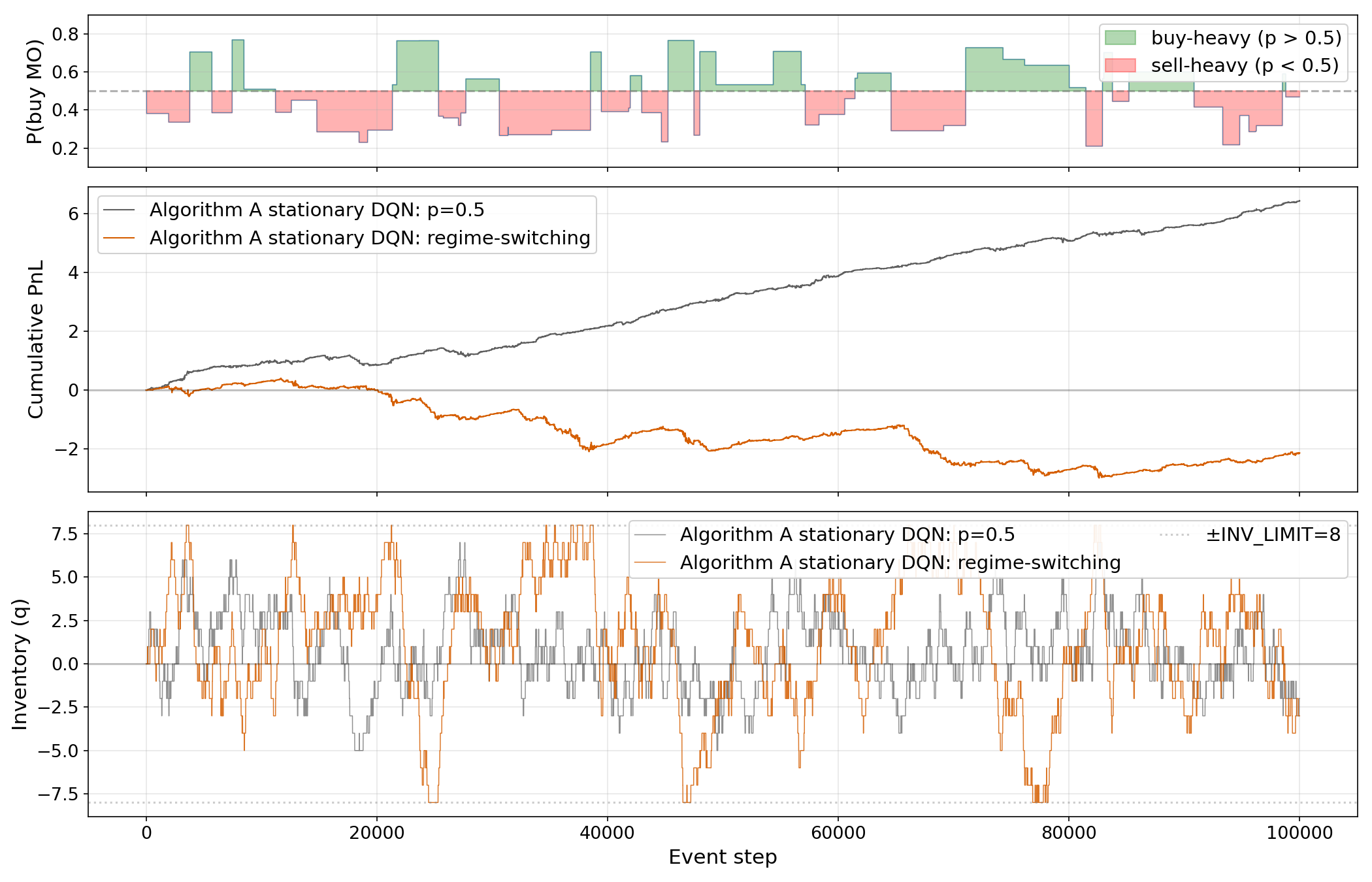}
    \caption{
    Single-episode diagnostic for Algorithm~A under regime-switching
    flow ($\tau_r=60$ MO events, $\omega=0.30$,
    $\varphi=10^{-3}$).
    Top: latent $p_{\mathrm{buy}}$ regimes
    (green: buy-heavy; red: sell-heavy).
    Middle: cumulative PnL under regime-switching (orange)
    and stationary symmetric flow (gray).
    Bottom: inventory trajectories, showing repeated
    saturation at the $\pm 8$ limits under regime-switching.
    }
  \label{fig:regime_episode}
\end{figure}

\paragraph{Single-episode diagnostic.}
We first examine how the stationarily trained Algorithm~A
responds to regime-switching order flow.
Figure~\ref{fig:regime_episode} illustrates a single
evaluation episode. The top panel shows the latent
$p_{\mathrm{buy}}$ schedule, with green and red regions
indicating buy- and sell-heavy regimes, respectively.
The middle panel compares the cumulative PnL of
Algorithm~A under regime-switching flow (orange)
and stationary symmetric flow (gray).
While the controller accumulates profits under symmetric
flow, persistent directional regimes produce substantial
drawdowns. The bottom panel shows the associated
inventory accumulation: the agent repeatedly reaches
the hard limits ($\pm 8$), with inventory saturation
coinciding with periods of deteriorating PnL.

\begin{figure}[htbp]
  \centering
  \includegraphics[width=0.7\linewidth]{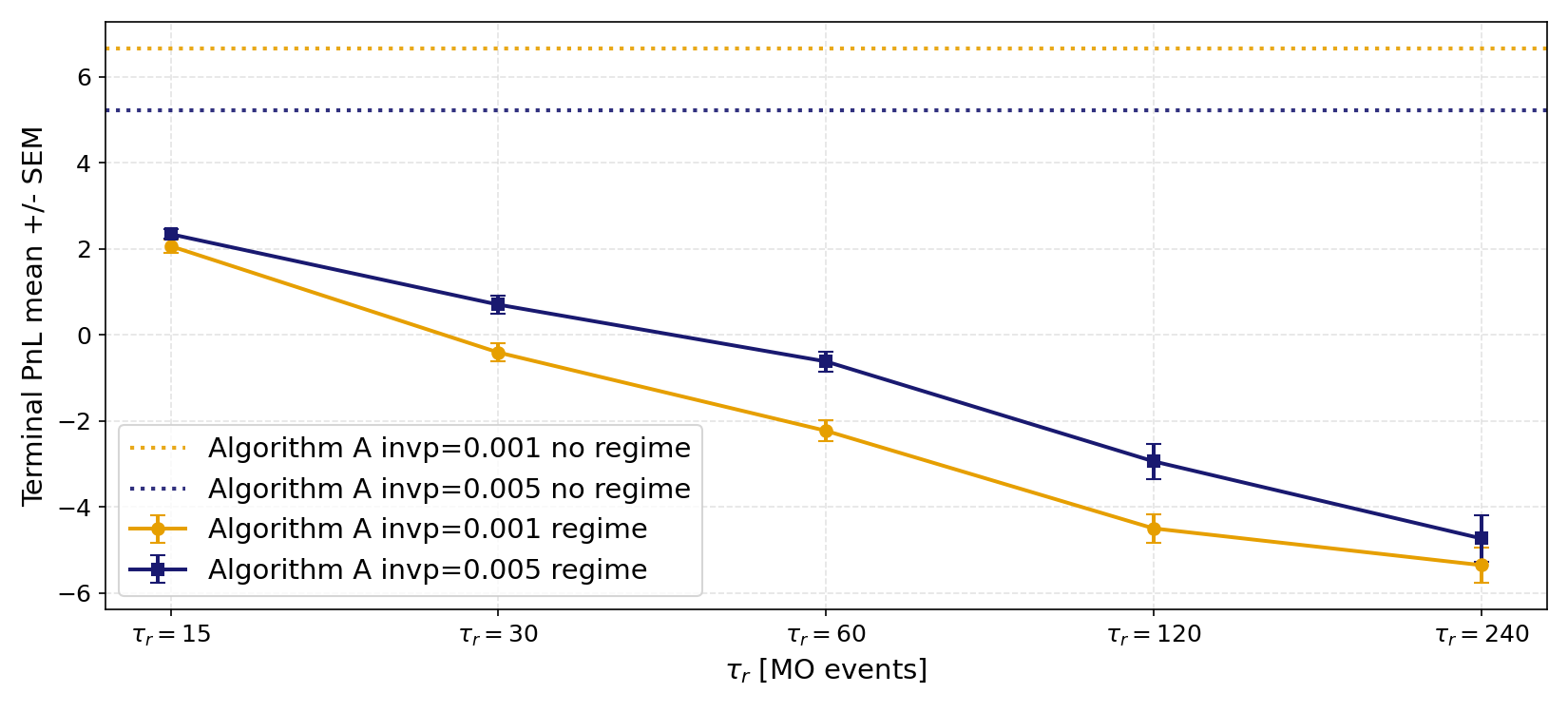}
    \caption{
    Mean terminal PnL of Algorithm~A under regime-switching flow
    versus regime persistence $\tau_r$ (MO clock), for
    $\varphi=10^{-3}$ (orange) and
    $\varphi=5\cdot10^{-3}$ (navy).
    Results use $N_{\mathrm{sim}}=100$ paired-seed episodes
    of $100\,000$ LOB events at $\omega=0.30$.
    Error bars show the standard error of the mean (SEM);
    dotted lines indicate the stationary-flow ($\omega=0$)
    baselines.
    }
  \label{fig:degradation_tau}
\end{figure}

\paragraph{Performance degradation as a function of $\tau_r$.}
Figure~\ref{fig:degradation_tau} extends the single-episode
diagnostic to $N_{\mathrm{sim}}=100$ paired-seed episodes,
evaluating two stationarily trained Algorithm~A controllers
($\varphi=10^{-3}$ and $\varphi=5\cdot10^{-3}$) across
$\tau_r\in\{15,30,60,120,240\}$ MO events, at fixed
$\omega=0.30$, against their stationary ($\omega=0$)
baselines. Mean terminal PnL decreases with regime
persistence for both controllers, consistent with the
GLFT results of Figure~\ref{fig:glft_degradation_tau}.
The controller with the weaker inventory penalty
($\varphi=10^{-3}$) eventually becomes loss-making,
while the stronger penalty ($\varphi=5\cdot10^{-3}$)
mitigates, but does not eliminate, the deterioration.
These results motivate the regime-aware training
introduced in Algorithm~B.

\section{Training under Regime-Switching Order Flow}
\label{sec:nonstat_training}

Algorithm~B is trained independently under the regime-switching
flow defined in Section~\ref{sec:robustness}, incorporating
the auxiliary features $e_t=(b_t,x_t)$ introduced in
Section~\ref{sec:learning_formulation}, which are absent
from Algorithm~A. We evaluate regime-aware training and
both signals jointly, keeping the action space, reward
function, and LOB simulator unchanged.

\paragraph{MO-clock implementation.}
Since regime durations are measured in market orders, the
regime schedule advances on MO arrivals rather than on every
LOB event. To pre-generate a regime schedule for each episode,
we estimate the number of MOs expected over
$N_{\mathrm{steps}}$ LOB events. Using
$E_{\mathrm{calib}}=5$ passive calibration episodes, we compute
\[
  \widehat p_{\mathrm{MO}}
  =
  \frac{1}{E_{\mathrm{calib}}}
  \sum_{e=1}^{E_{\mathrm{calib}}}
  \frac{N^{\mathrm{MO}}_e}{N_{\mathrm{steps}}},
\]
where $N^{\mathrm{MO}}_e$ is the MO count in episode $e$.
The estimated MO horizon, rounded to the nearest integer, is
\begin{equation}
  \widehat M
  =
  \left\lfloor
    N_{\mathrm{steps}}\,\widehat p_{\mathrm{MO}}+\frac12
  \right\rfloor.
  \label{eq:mo_horizon}
\end{equation}
A buffered regime schedule is generated before each episode,
independently of the MM's actions. During simulation, both
the regime schedule and the Bayesian filter advance only
on MO arrivals, using the same MO-event clock.

\begin{figure}[htbp]
  \centering
  \includegraphics[width=0.8\linewidth]{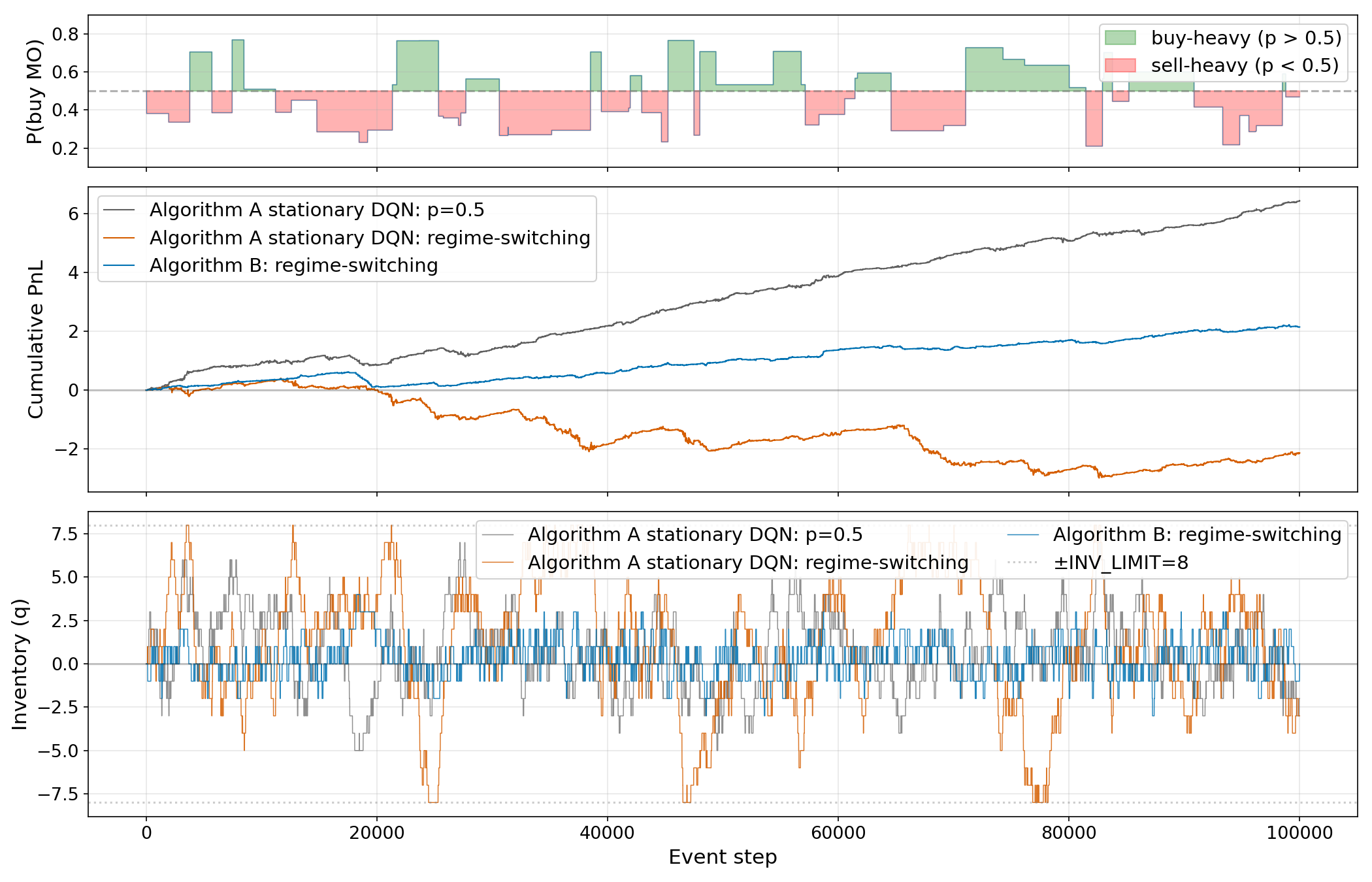}
  \caption{
  Single-episode comparison under regime-switching flow
  ($\tau_r=60$ MO events, $\omega=0.30$).
  Top: latent $p_{\mathrm{buy}}$ regimes
  (green: buy-heavy; red: sell-heavy).
  Middle: cumulative PnL of Algorithm~A under symmetric flow
  (gray), Algorithm~A under regime-switching (orange),
  and Algorithm~B under regime-switching (blue).
  Bottom: corresponding inventory trajectories, with
  hard limits at $\pm q_{\max}$.
  }
  \label{fig:regime_trained}
\end{figure}

\paragraph{Algorithm~B result.}
Figure~\ref{fig:regime_trained} revisits the episode of
Figure~\ref{fig:regime_episode}, adding Algorithm~B trained
under regime-switching flow ($\tau_r=60$ MO events,
$\omega=0.30$). Unlike the stationarily trained Algorithm~A,
which incurs substantial losses under directional regimes,
Algorithm~B achieves positive cumulative PnL while
maintaining a tighter inventory trajectory, rarely reaching
the hard limits $\pm q_{\max}$. This suggests that
regime-aware training improves inventory control and
mitigates the losses observed under persistent directional flow.

\begin{figure}[htbp]
  \centering
  \includegraphics[width=0.7\linewidth]{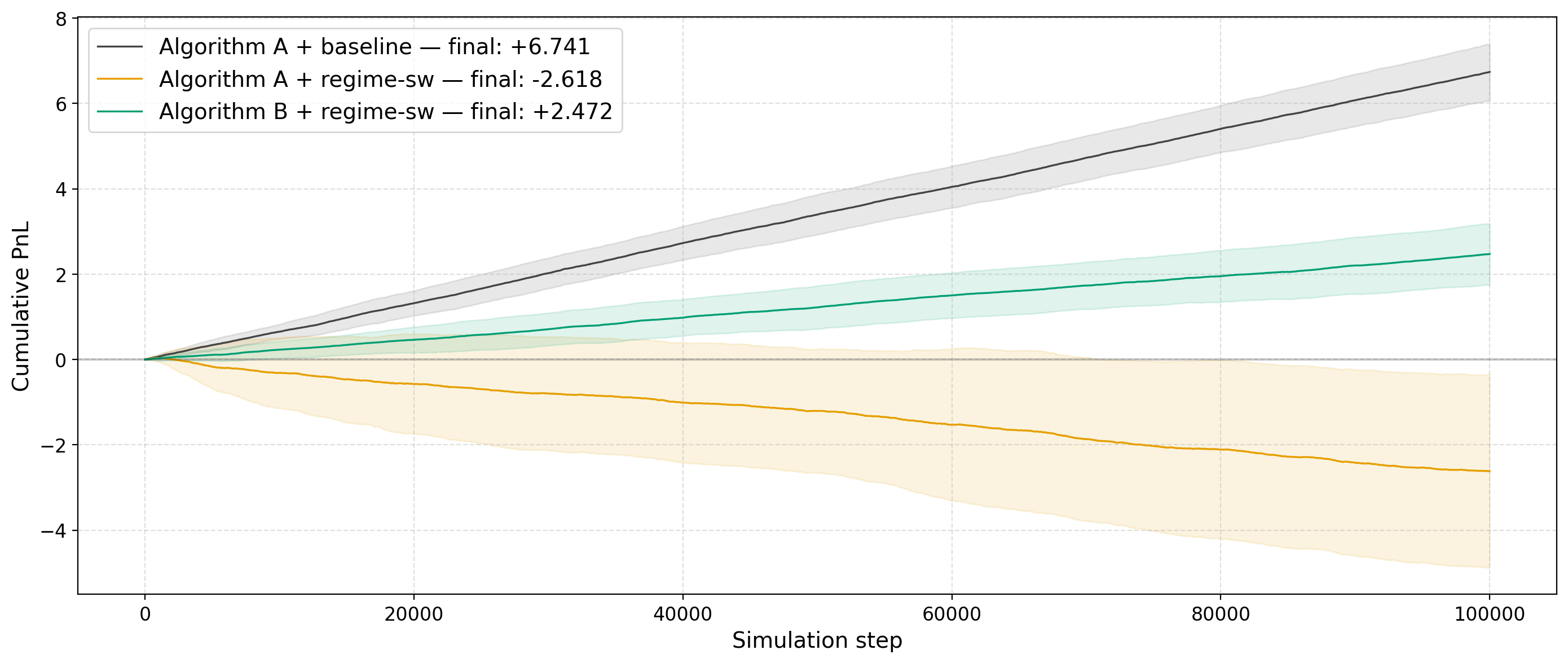}
  \caption{
  Mean cumulative PnL $\pm 1$ SD over $100$ paired-seed
  episodes of $100\,000$ LOB events
  ($\tau_r=60$, $\omega=0.30$).
  Algorithm~A under symmetric flow (gray),
  Algorithm~A under regime-switching (orange),
  and Algorithm~B under regime-switching (green).
  }
  \label{fig:regime_trained_mean}
\end{figure}

\paragraph{Monte Carlo simulations.}
We evaluate the three policies over $N_{\mathrm{sim}}=100$
paired-seed episodes of $N_{\mathrm{steps}}=100\,000$
LOB events, with $\tau_r=60$ MO events and $\omega=0.30$.
Figure~\ref{fig:regime_trained_mean} shows mean cumulative
PnL $\pm 1$ standard deviation (SD) across episodes.
Under regime-switching flow, Algorithm~A becomes
loss-making, whereas Algorithm~B restores positive mean
PnL and recovers a substantial fraction of the stationary
benchmark's profitability. The improvement is statistically
significant under a paired $t$-test ($p<10^{-4}$).

\begin{figure}[htbp]
  \centering
  \includegraphics[width=0.7\linewidth]{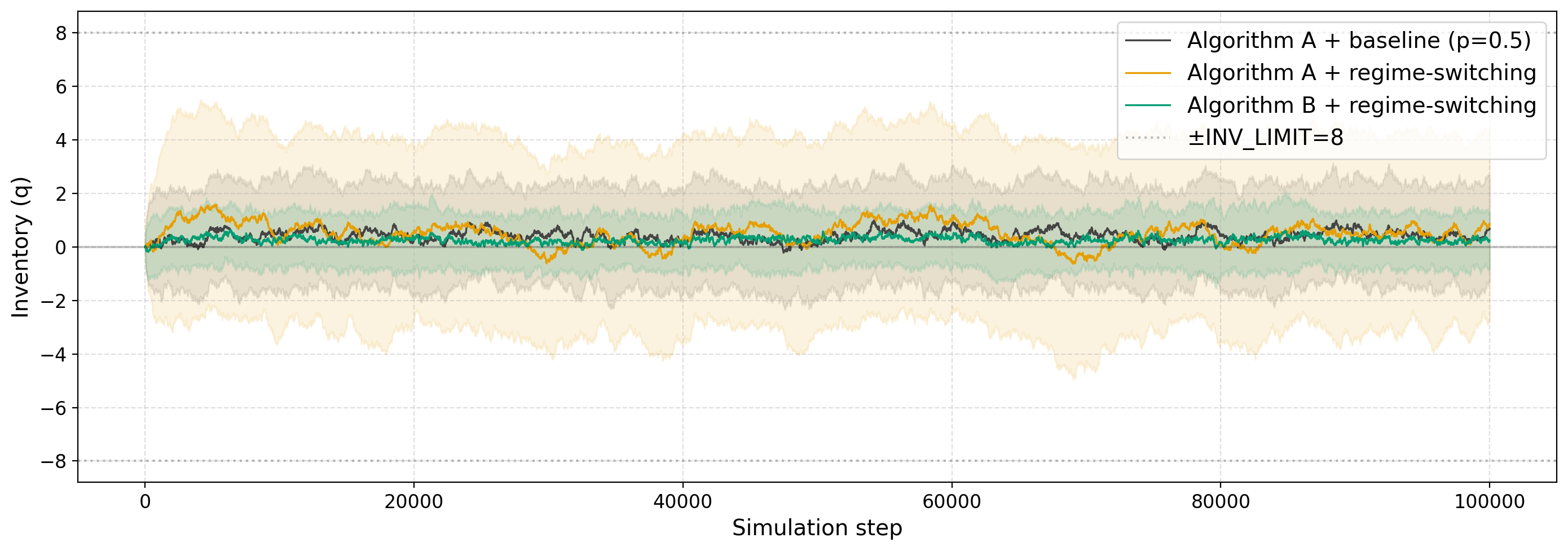}
  \caption{
  Inventory trajectories (mean $\pm 1$ SD) over the same
  $100$ paired-seed episodes.
  Algorithm~A under symmetric flow (gray),
  Algorithm~A under regime-switching (orange),
  and Algorithm~B under regime-switching (green).
  Dashed lines indicate the inventory limits
  $\pm q_{\max}=\pm 8$.
  }
  \label{fig:regime_trained_inv}
\end{figure}

\paragraph{Inventory control.}
Figure~\ref{fig:regime_trained_inv} shows the corresponding
inventory trajectories. Under regime-switching flow,
Algorithm~A exhibits substantially greater inventory
dispersion, whereas Algorithm~B maintains even tighter
inventory control than the stationary-flow benchmark.
This reduction in inventory exposure accompanies the
recovery in PnL.

\begin{figure}[htbp]
  \centering
  \includegraphics[width=0.7\linewidth]{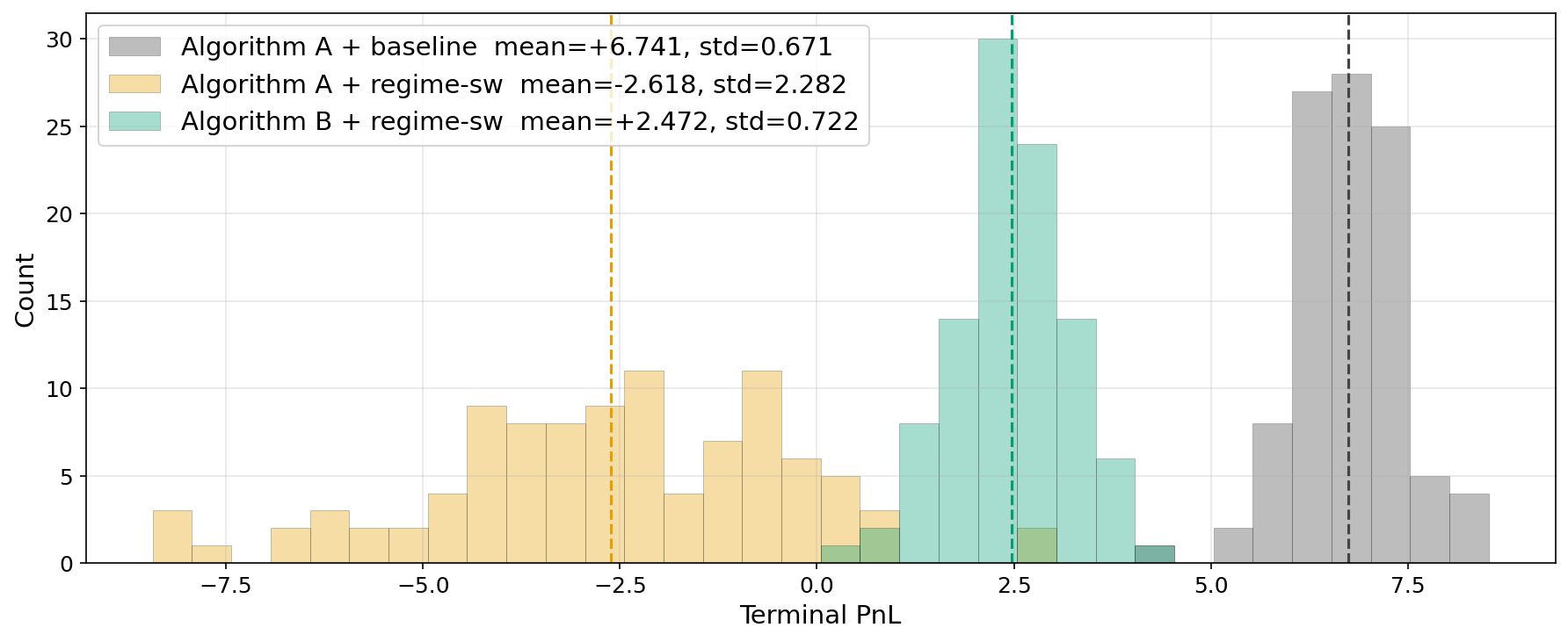}
  \caption{
  Terminal-PnL distributions over $N_{\mathrm{sim}}=100$
  paired-seed episodes of $100\,000$ LOB events.
  Algorithm~A under symmetric flow (gray),
  Algorithm~A under regime-switching (orange), and
  Algorithm~B under regime-switching (green).
  Mean PnL and SD, normalized by the symmetric-flow
  Algorithm~A mean, are $(100\%,10\%)$,
  $(-39\%,34\%)$, and $(37\%,11\%)$, respectively.
  Dashed lines indicate the sample means.
  }
  \label{fig:pnl_distribution}
\end{figure}

\paragraph{Terminal-PnL distributions.}
Figure~\ref{fig:pnl_distribution} shows the terminal-PnL
distributions across the same $100$ episodes.
Under regime-switching flow, Algorithm~A's distribution
shifts toward losses and becomes substantially more
dispersed. In contrast, Algorithm~B restores positive
mean PnL with a dispersion comparable to the
stationary-flow benchmark.

\section{Interpreting the Regime-Trained Policy}
\label{sec:interpretability}

We examine how Algorithm~B's learned value function
depends on inventory, the bid--ask spread, and inferred
directional flow, assessing whether these relationships
are consistent with market-making intuition.
The quote-exposure signal is examined separately below.

We evaluate the network on two-dimensional slices of a
reference state, varying two features while holding the
remaining coordinates fixed.
\footnote{The reference state has a one-tick spread,
zero inventory, four units at each observed depth level,
and zero directional bias and quote-exposure imbalance.
The regime-age feature is fixed at its normalized value
of $0.7$ (Appendix~\ref{app:bayesian_flow_filter}).}

At each state, we compute
\begin{equation}
  \widehat V_\theta(s)
  =
  \max_{a\in\mathcal A(s)} Q_\theta(s,a),
\end{equation}
where $\mathcal A(s)$ denotes the admissible actions under
the inventory constraint, and $Q_\theta$ is the expected
C51 return defined in Section~\ref{sec:learning_formulation}.
The resulting heatmaps characterize local variations
in the learned value function.

\begin{figure}[htbp]
  \centering
  \makebox[\linewidth][c]{%
  \begin{minipage}{1.18\linewidth}
  \centering
  \begin{subfigure}[t]{0.333\linewidth}
    \centering
    \includegraphics[width=\linewidth]{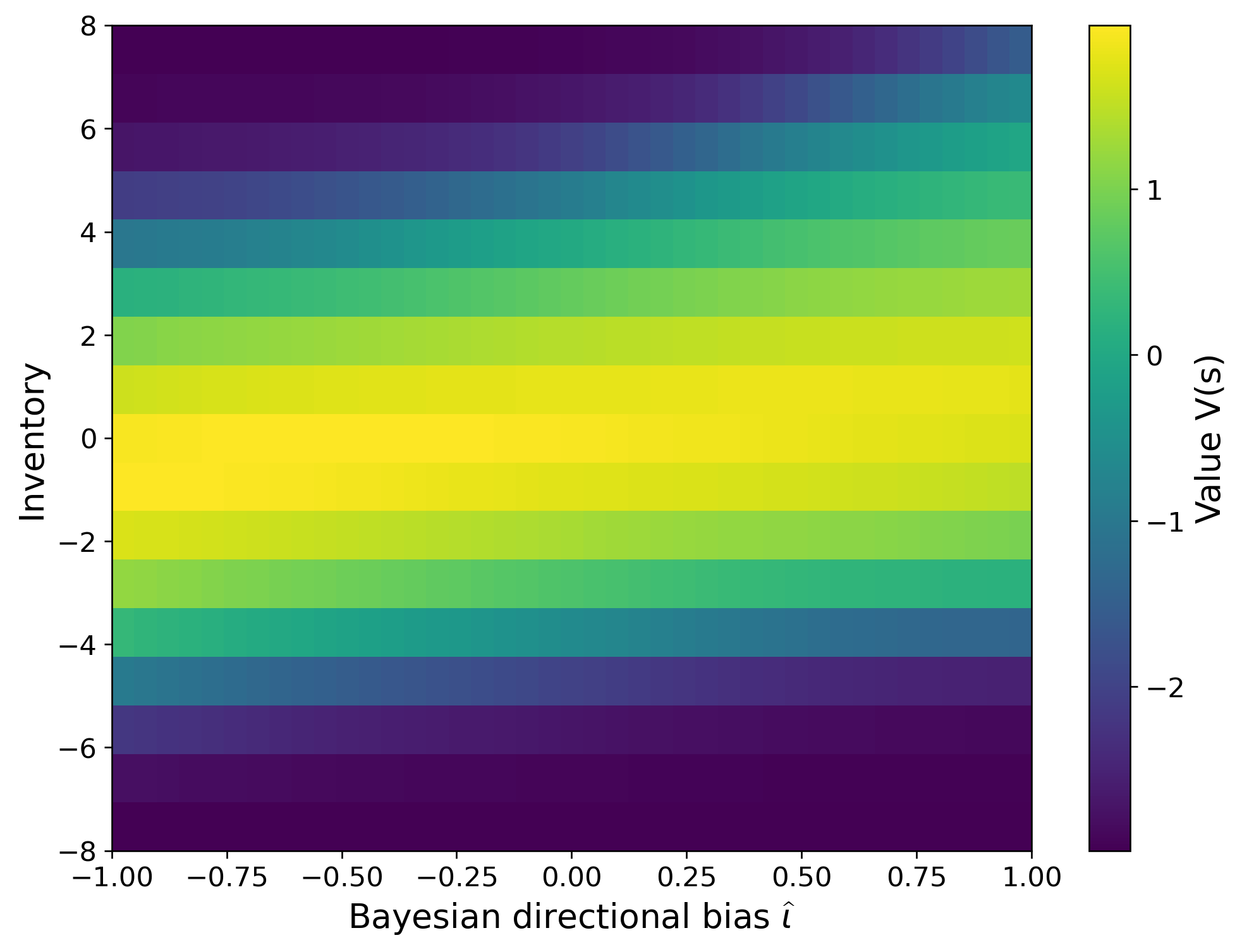}
    \caption{Inventory and Bayesian bias.}
    \label{fig:interp_heat_inv_bayes}
  \end{subfigure}\hfill
  \begin{subfigure}[t]{0.333\linewidth}
    \centering
    \includegraphics[width=\linewidth]{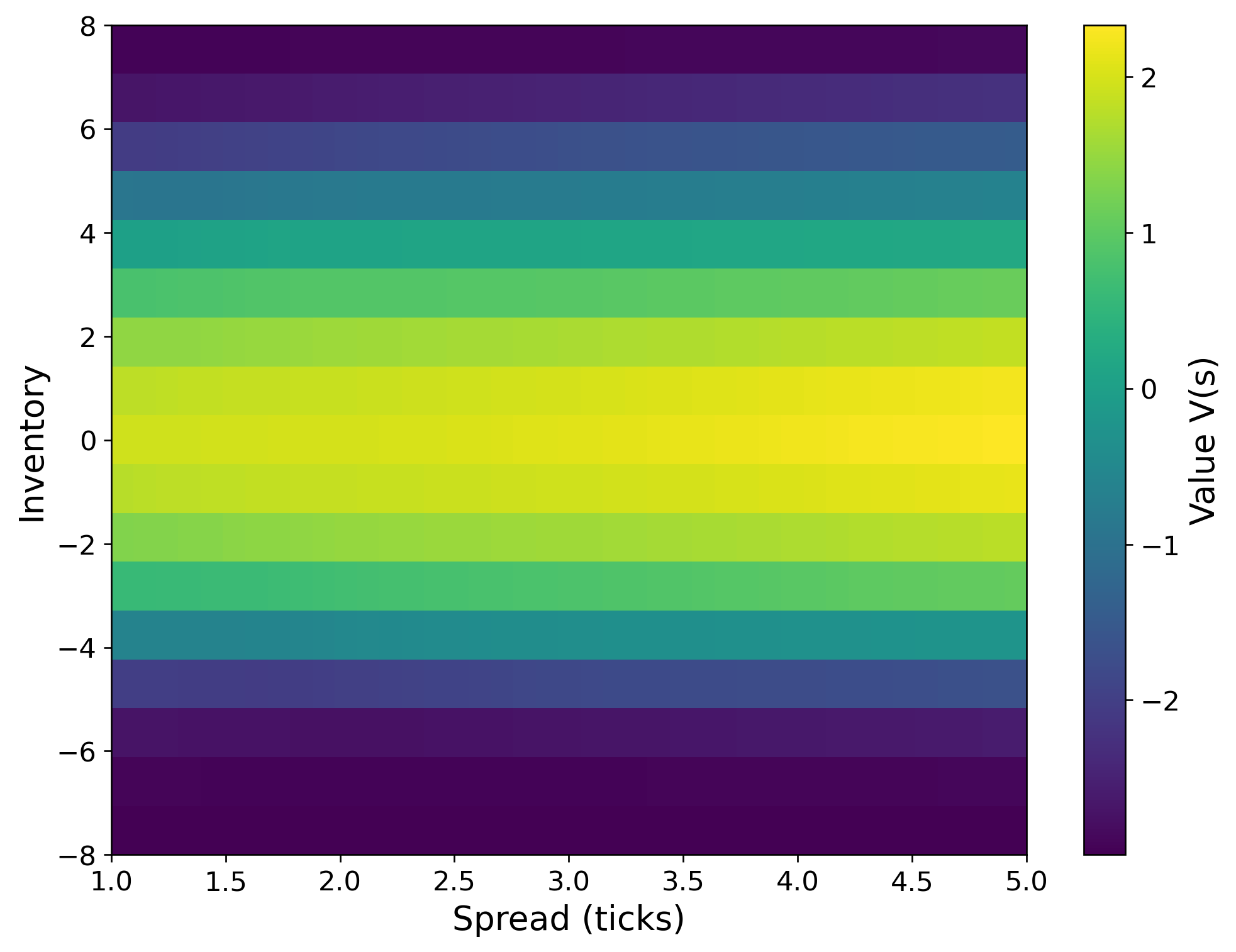}
    \caption{Inventory and spread.}
    \label{fig:interp_heat_inv_spread}
  \end{subfigure}\hfill
  \begin{subfigure}[t]{0.333\linewidth}
    \centering
    \includegraphics[width=\linewidth]{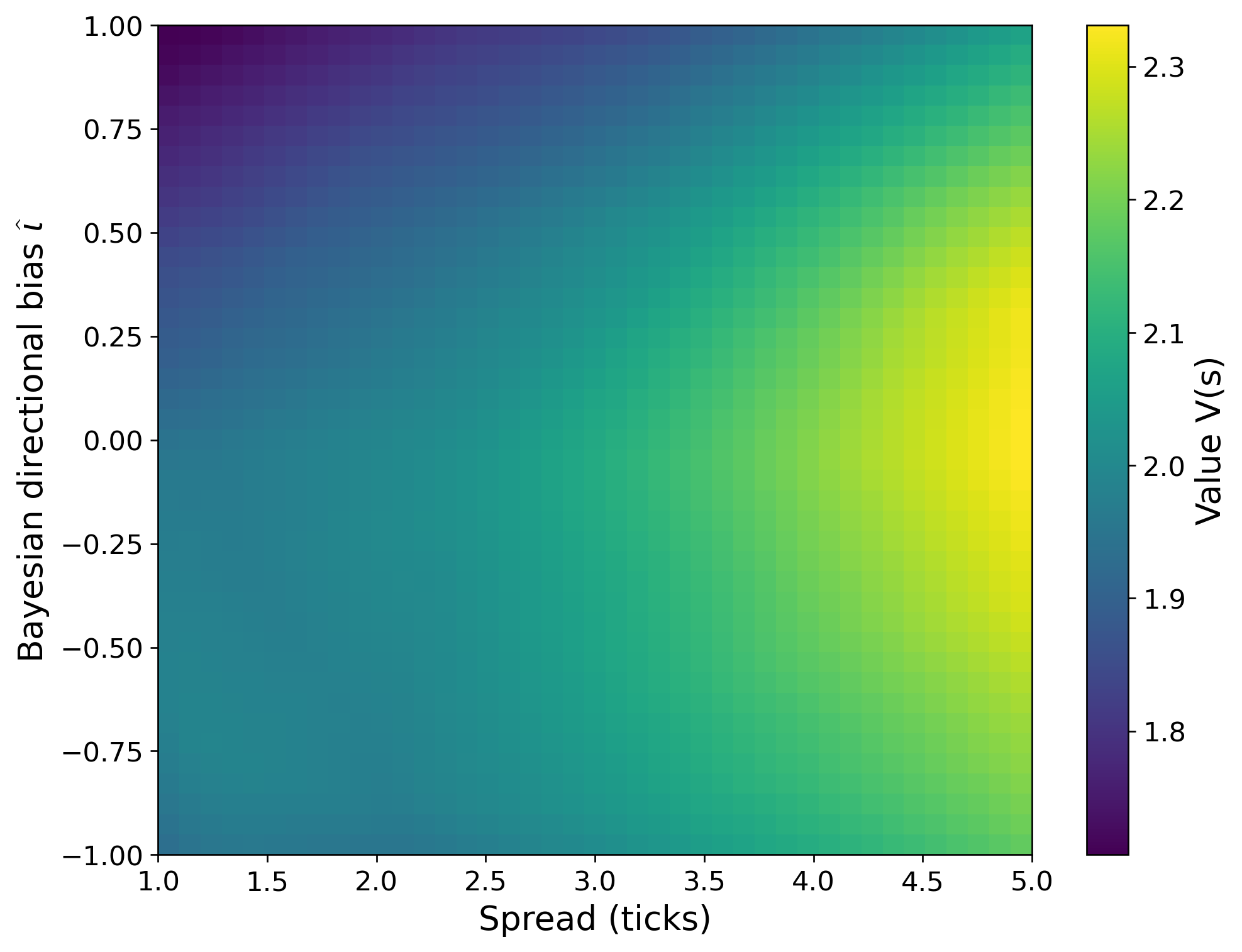}
    \caption{Bayesian bias and spread.}
    \label{fig:interp_heat_bayes_spread}
  \end{subfigure}
  \end{minipage}}
    \caption{
    Two-dimensional slices of Algorithm~B's learned value
    function, varying two state features while holding the
    remaining ones fixed at their reference values.
    }
  \label{fig:interp_heatmaps}
\end{figure}

\paragraph{Heatmap slices.}
Figure~\ref{fig:interp_heatmaps} shows three slices of
Algorithm~B's learned value function.
The inventory--regime slice
(Figure~\ref{fig:interp_heatmaps}\subref{fig:interp_heat_inv_bayes})
shows higher values near zero inventory and a marked
decline near the hard limits. For nonzero inventory,
the Bayesian belief shifts the value surface: long
positions are assigned higher value under buy-heavy
flow, which favors ask executions, while the opposite
holds for short positions under sell-heavy flow.

The inventory--spread slice
(Figure~\ref{fig:interp_heatmaps}\subref{fig:interp_heat_inv_spread})
shows that wider spreads increase value primarily
at moderate inventory levels. Similarly, the
regime--spread slice
(Figure~\ref{fig:interp_heatmaps}\subref{fig:interp_heat_bayes_spread})
shows that the value of wider spreads diminishes under
strong directional imbalances.
Together, these patterns are consistent with the
trade-off between spread capture and inventory risk,
within the reference-state neighborhood examined.

\begin{figure}[htbp]
  \centering
  \begin{subfigure}[b]{0.7\linewidth}
    \centering
    \includegraphics[width=\linewidth]{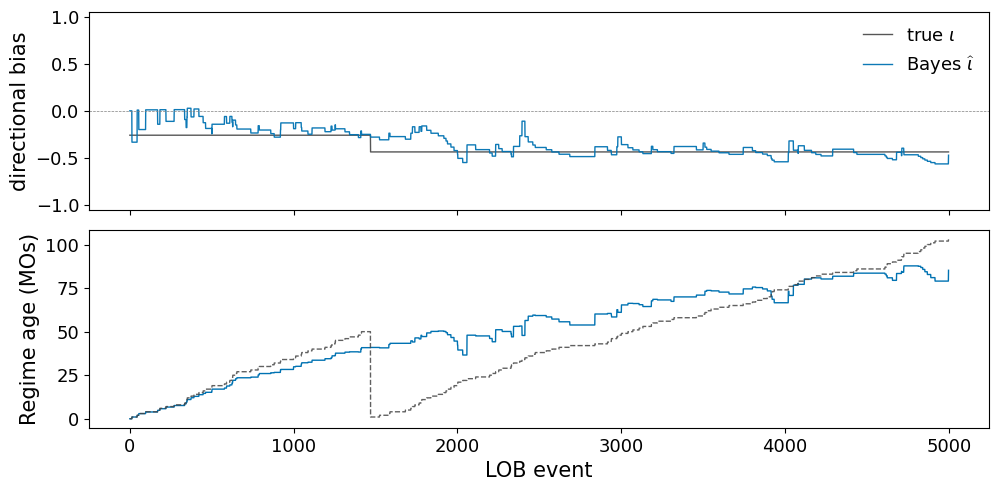}
    \caption{Bayesian regime-belief diagnostics.}
    \label{fig:interp_bayes_example}
  \end{subfigure}

  \vspace{6pt}

  \begin{subfigure}[b]{0.7\linewidth}
    \centering
    \includegraphics[width=\linewidth]{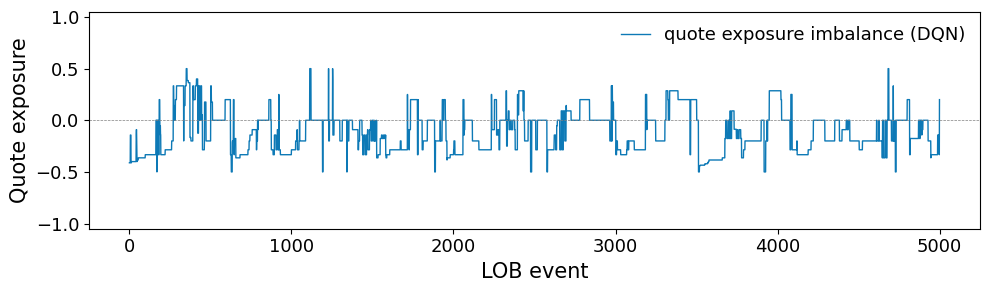}
    \caption{Quote-exposure imbalance diagnostic.}
    \label{fig:interp_quote_exposure_example}
  \end{subfigure}
    \caption{
    Auxiliary signals during a $5\,000$-event evaluation
    episode of Algorithm~B with frozen weights:
    Bayesian directional bias, posterior expected regime age,
    and quote-exposure imbalance.
    }
  \label{fig:interp_signal_examples}
\end{figure}

\paragraph{Auxiliary-signal diagnostics.}
Figure~\ref{fig:interp_signal_examples} illustrates the
auxiliary signals during an evaluation episode.
The Bayesian estimate $\widehat{\iota}_t$ tracks changes
in the latent directional bias, while the posterior
expected run length increases within regimes and
declines around detected transitions.
The quote-exposure signal $x_t$ captures fluctuations
in the relative fill exposure of the MM's resting orders,
complementing the information provided by the
Bayesian filter.

\section{Out-of-Distribution Regime Stress Tests}
\label{sec:stress_tests}

The preceding evaluation uses the training regime family
($\tau_r=60$ MO events, $p_{\mathrm{buy}}\in[0.20,0.80]$).
We examine Algorithm~B's robustness to changes in
regime persistence and directional dependence through
three stress tests with paired random seeds.
The first uses $100\,000$ LOB events per episode,
while the remaining tests use $500\,000$; absolute
PnL levels are therefore compared only within each test.

\begin{figure}[htbp]
  \centering
  \includegraphics[width=0.7\linewidth]{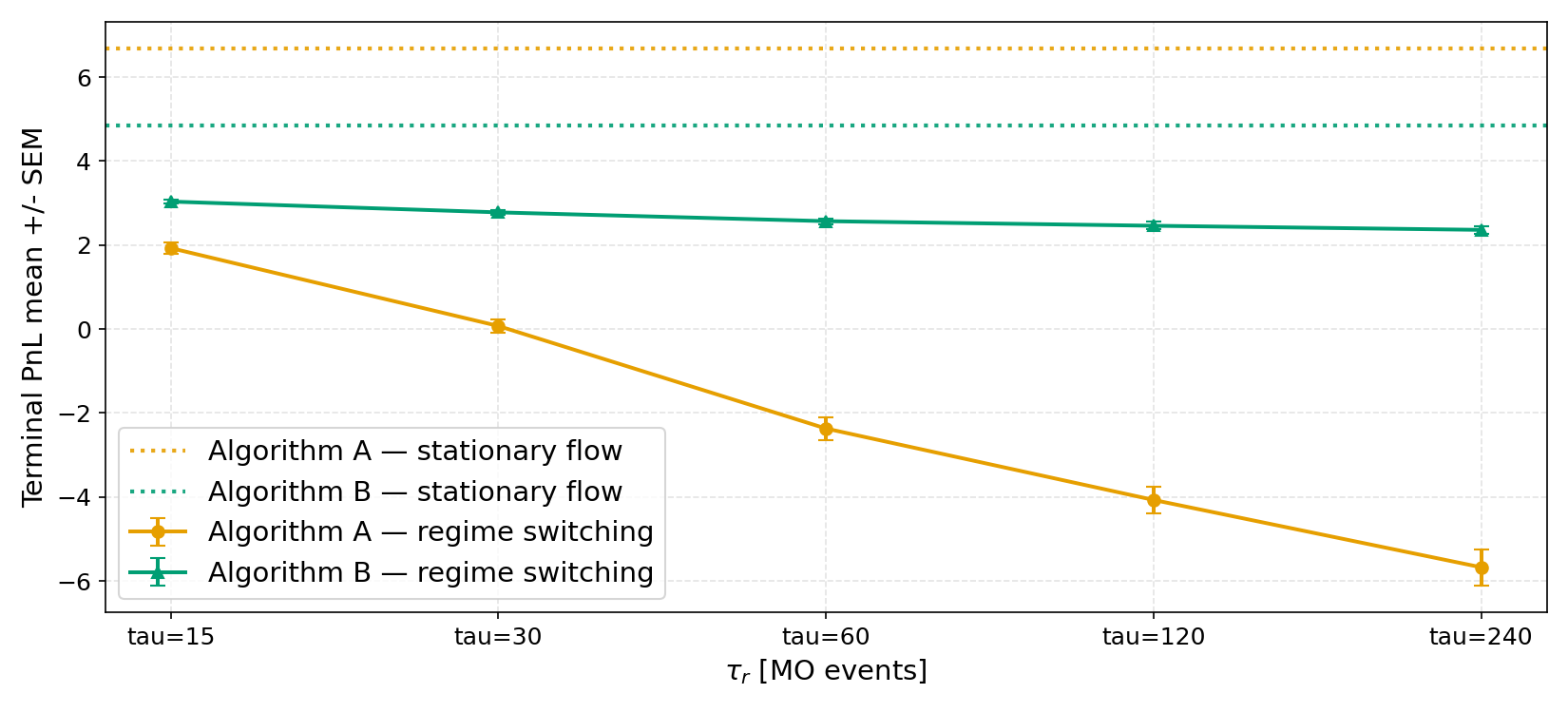}
  \caption{
  Mean terminal PnL $\pm$ SEM for Algorithms~A and B
  across $\tau_r\in\{15,30,60,120,240\}$ MO events,
  with $p_{\mathrm{buy}}\in[0.20,0.80]$ and
  $\varphi=10^{-3}$.
  Results use $100$ paired-seed episodes of
  $100\,000$ LOB events.
  Horizontal lines indicate each controller's
  stationary-flow benchmark.
  }
  \label{fig:phaseb_tau_sweep}
\end{figure}

\paragraph{Persistence sweep.}
We vary the regime-duration scale over
$\tau_r\in\{15,30,60,120,240\}$ MO events, keeping
$p_{\mathrm{buy}}\in[0.20,0.80]$.
Figure~\ref{fig:phaseb_tau_sweep} compares Algorithms~A
and B over $100$ paired episodes of $100\,000$ LOB
events. As regime persistence increases, Algorithm~A
exhibits progressively lower mean PnL and eventually
becomes loss-making. Algorithm~B, in contrast, remains
profitable across all persistence scales, despite some
deterioration under longer-lived directional flow.

\begin{figure}[htbp]
  \centering
  \begin{subfigure}[t]{0.48\textwidth}
    \includegraphics[width=\linewidth]{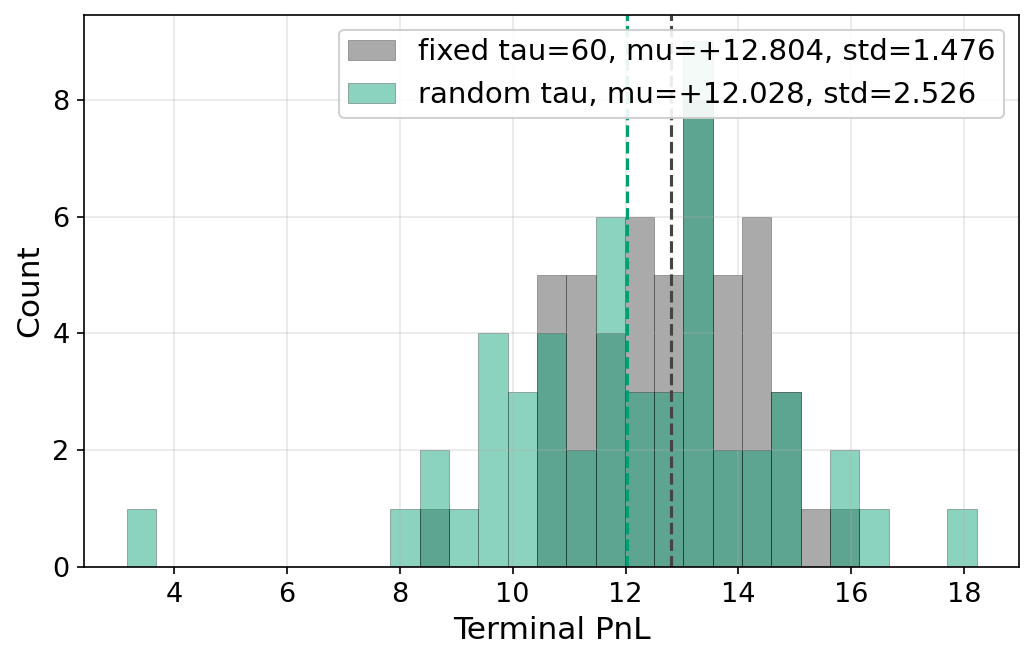}
    \caption{Random regime persistence.}
    \label{fig:random_tau_stress}
  \end{subfigure}
  \hfill
  \begin{subfigure}[t]{0.48\textwidth}
    \includegraphics[width=\linewidth]{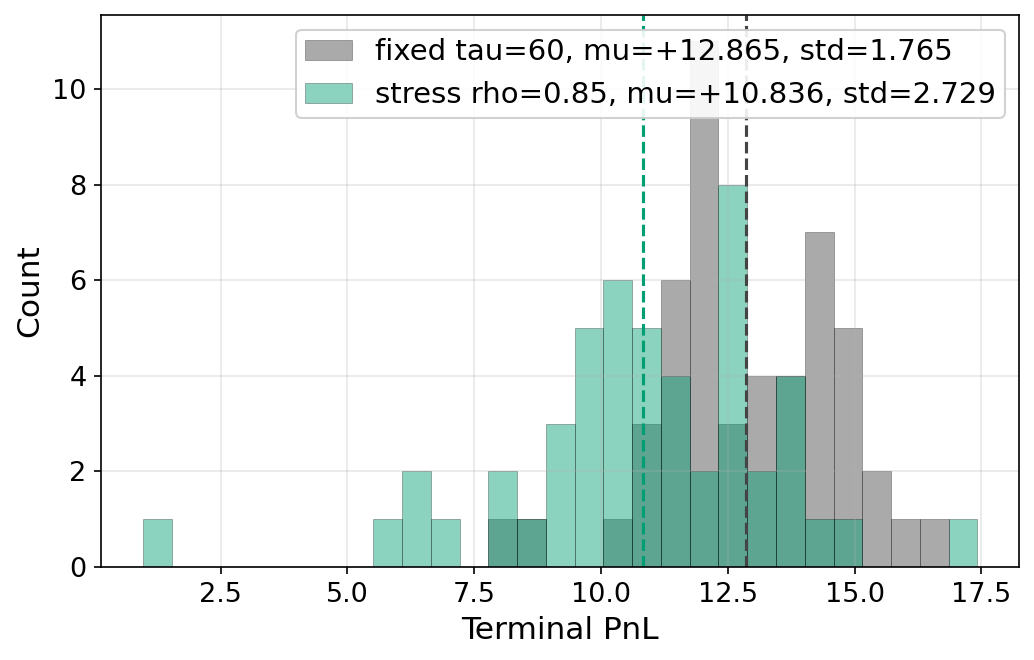}
    \caption{Correlated directional regimes.}
    \label{fig:correlated_regime_stress}
  \end{subfigure}
  \caption{
  Terminal-PnL distributions of Algorithm~B under
  (a) random regime persistence and
  (b) additional directional correlation
  ($\rho_{\mathrm{side}}=0.85$), compared with the
  fixed-$\tau_r=60$ benchmark.
  Each test uses $50$ paired-seed episodes of
  $500\,000$ LOB events.
  Relative to the benchmark, mean terminal PnL
  decreases by $6.1\%$ and $15.8\%$, respectively.
  }
  \label{fig:phaseB_stress_combined}
\end{figure}

\paragraph{Random regime persistence.}
We next allow the persistence scale to vary between
regimes, drawing
$\tau_k\sim\mathrm{Uniform}\{15,30,60,120,240\}$
and $L_k=\lceil X_k\rceil$, where
$X_k\sim\mathrm{Exp}(1/\tau_k)$.
The buy-probability range remains unchanged, and
$\tau_k$ is hidden from the controller.
Figure~\ref{fig:random_tau_stress} compares Algorithm~B
against its fixed-$\tau_r=60$ benchmark over $50$
paired episodes of $500\,000$ events.
Mean PnL remains positive, although the distribution
becomes more dispersed, with lower outcomes in its
left tail.

\paragraph{Correlated directional regimes.}
Finally, we introduce directional dependence into the
random-persistence model. At each regime boundary,
the directional sign $s_k\in\{-1,+1\}$ is retained
with probability $\rho_{\mathrm{side}}=0.85$, while
\[
  I_k\sim\mathrm{Uniform}[0,0.30],
  \qquad
  p_{\mathrm{buy},k}=0.50+s_k I_k.
\]
This preserves the buy-probability range but allows
directional pressure to persist across regime boundaries.
Algorithm~B remains profitable
(Figure~\ref{fig:correlated_regime_stress}), although
its mean PnL declines further and the distribution
exhibits greater downside dispersion.

These results indicate that Algorithm~B maintains
positive mean PnL beyond its training regime family,
although greater persistence and directional dependence
increase downside risk.

\section{Algorithm C: Scenario-Bandit Fine-Tuning}
\label{sec:scenario_bandit_finetuning}

The stress tests in Section~\ref{sec:stress_tests} show that
Algorithm~B remains profitable beyond its training regime family,
but exhibits greater downside dispersion. We introduce Algorithm~C,
which fine-tunes Algorithm~B by allocating more training episodes
to regime scenarios associated with poor performance.

\paragraph{Scenario generation.}
A scenario $\xi_i$ specifies an exogenous sequence of regimes
on the market-order clock:
\[
  \xi_i=
  \bigl((\tau_{i,k},L_{i,k},p_{i,k})_{k=1}^{K_i},\zeta_i\bigr),
\]
where $K_i$ is the number of regimes, $\tau_{i,k}$ their duration
scales, $L_{i,k}$ their realized durations, $p_{i,k}$ their buy-MO
probabilities, and $\zeta_i$ is the random seed used to generate the scenario.
Each sequence is generated to cover the estimated
MO horizon $\widehat M$ of Eq.~\eqref{eq:mo_horizon}.

The initial pool contains $M=256$ scenarios, equally
divided between the random-persistence and
correlated-direction models of Section~\ref{sec:stress_tests}.
During fine-tuning, replacement scenarios are drawn
from either model with equal probability, defining
the reference distribution $P_0$.

\paragraph{Adaptive scenario sampling.}
The mechanism is related to Prioritized Level Replay
\citep{jiang2021prioritized}, which prioritizes training
environments according to their estimated learning potential.
Here, the bandit assigns higher sampling probabilities
to scenarios associated with a lower inventory-adjusted
terminal PnL.
This scenario-level selection complements the
transition-level prioritized experience replay used
by the C51 learner.

Let $t=0,\ldots,T-1$ index the decision intervals in an
episode, with $n=0,\ldots,N_t-1$ indexing the LOB events
within interval $t$, as in
Section~\ref{sec:learning_formulation}. After evaluating the
policy $\pi_\theta$ under scenario $\xi_i$, we compute
\[
  H_T(\theta;\xi_i)
  =
  \mathrm{PnL}_T(\theta;\xi_i)
  -
  \varphi_{\mathrm{sb}}\sum_{t=0}^{T-1}\sum_{n=0}^{N_t-1}
  \left[
    (q^+_{t,n})^2
    +
    (|q^+_{t,n}|-q_w)_+^2
  \right],
\]
where $\mathrm{PnL}_T(\theta;\xi_i)$ denotes terminal PnL
and $q^+_{t,n}$ is the post-event inventory, as in
Eq.~\eqref{eq:reward}. The score uses a monetary-scale penalty
coefficient $\varphi_{\mathrm{sb}}=10^{-3}$, while the learning
reward is expressed in ticks. It is undiscounted and includes
full terminal PnL. It determines scenario sampling;
the discounted C51 rewards continue to govern policy updates.

Defining the scenario loss as
$\ell_i(\theta)=-H_T(\theta;\xi_i)$, we update the difficulty
of the scenario $i$, when selected in the fine-tuning episode $u$ with
network parameters $\theta_u$, according to
\begin{equation}
  d_i^{(u+1)}
  =
  (1-\eta_{\mathrm{ewma}})d_i^{(u)}
  +
  \eta_{\mathrm{ewma}}\ell_i(\theta_u),
  \qquad
  \eta_{\mathrm{ewma}}=0.05,
  \label{eq:difficulty_ewma}
\end{equation}
with $d_i^{(0)}=0$. Non-selected scenarios retain their scores.

To favor difficult scenarios, we define
\[
  \begin{aligned}
  \nu_i
  &=
  \frac{d_i-\bar d}{s_d+\epsilon_\nu},
  \\[3pt]
  \widetilde w_i
  &=
  \frac{\exp(\beta_{\mathrm{sb}}\nu_i)}
       {\sum_{j=1}^{M}\exp(\beta_{\mathrm{sb}}\nu_j)},
  \\[3pt]
  w_i
  &=
  (1-\varepsilon)\widetilde w_i
  +\frac{\varepsilon}{M},
  \end{aligned}
\]
where $\bar d$ and $s_d$ are the mean and standard deviation
of the pool difficulties, and $\epsilon_\nu>0$ is a small
constant to avoid division by zero. If $s_d$ is numerically zero, we set all $\nu_i=0$.
The parameter $\beta_{\mathrm{sb}}$ controls the emphasis on
difficult scenarios, while $\varepsilon$ mixes the resulting
weights with uniform sampling.

To prevent excessive concentration on individual
scenarios, sampling probabilities are capped at
$w_{\max}=0.05$. Excess probability is redistributed
among scenarios below this limit, proportionally to
their remaining capacity,
\[
  \Delta w_j \propto w_{\max}-w_j,
  \qquad w_j<w_{\max},
\]
while preserving $\sum_{j=1}^{M}w_j=1$.

During the first
half of fine-tuning, $\beta_{\mathrm{sb}}$ increases linearly
from $0$ to $0.5$ and $\varepsilon$ decreases from $1$ to
$0.5$. Every $500$ bandit updates, the easiest $10\%$ of
scenarios are replaced by new draws from $P_0$, initialized
at the pool's median difficulty.

\paragraph{Policy fine-tuning.}
Algorithm~C starts from Algorithm~B, retaining its state and
action spaces and distributional C51 learning rule.
To mitigate the risk of catastrophic forgetting, we
penalize deviations from the pretrained parameters:
\[
  \mathcal L_C(\theta)
  =
  \mathcal L_{\mathrm{C51}}(\theta)
  +
  \lambda_{\mathrm{anchor}}\|\theta-\theta_B\|_2^2,
  \qquad
  \lambda_{\mathrm{anchor}}=10^{-3},
\]
where $\theta_B$ denotes the fixed parameters of Algorithm~B.

\begin{figure}[htbp]
  \centering
  \includegraphics[width=0.5\linewidth]{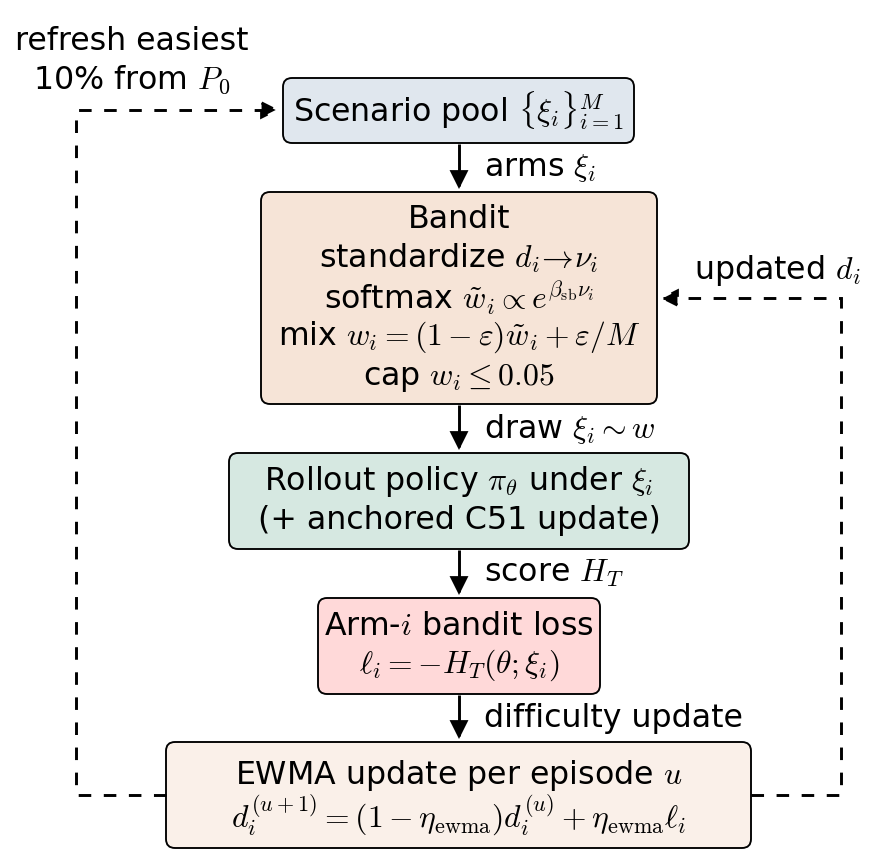}
    \caption{
    Scenario-bandit fine-tuning.
    A scenario is sampled from the current pool according
    to difficulty-based bandit weights, used for policy
    rollout and anchored C51 updates, and then assigned
    an updated difficulty score. The easiest scenarios
    are periodically replaced by fresh draws from $P_0$.
    }
  \label{fig:scenario_bandit_loop}
\end{figure}

Fine-tuning is performed over $2\,000$ episodes of
$5\,000$ LOB events, using the adaptive scenario
sampling procedure described above.
Figure~\ref{fig:scenario_bandit_loop}
summarizes the method.

\begin{figure}[htbp]
  \centering
  \begin{subfigure}[t]{0.48\textwidth}
    \includegraphics[width=\linewidth]{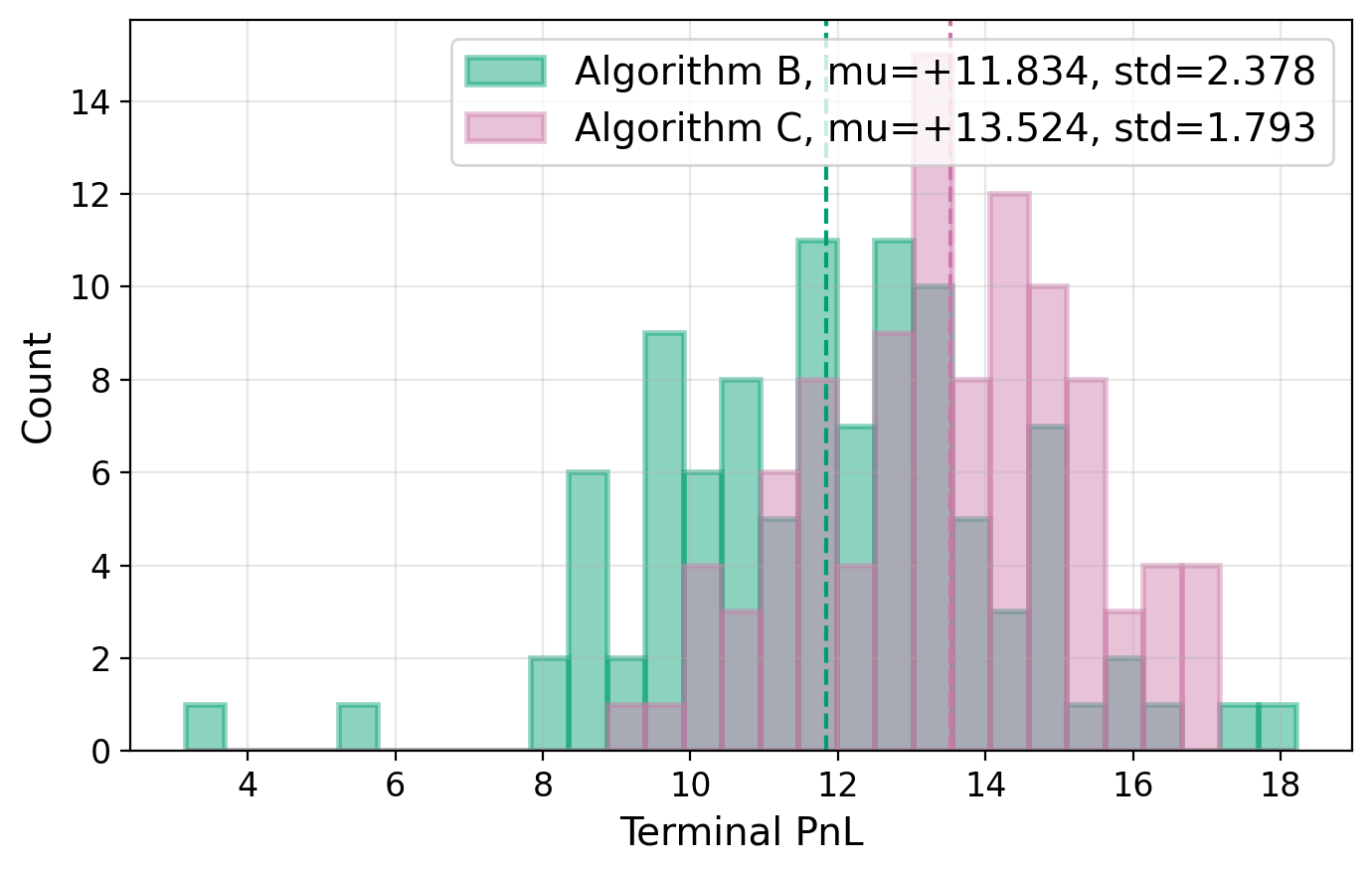}
    \caption{Random regime persistence.}
    \label{fig:bandit_random_tau}
  \end{subfigure}
  \hfill
  \begin{subfigure}[t]{0.48\textwidth}
    \includegraphics[width=\linewidth]{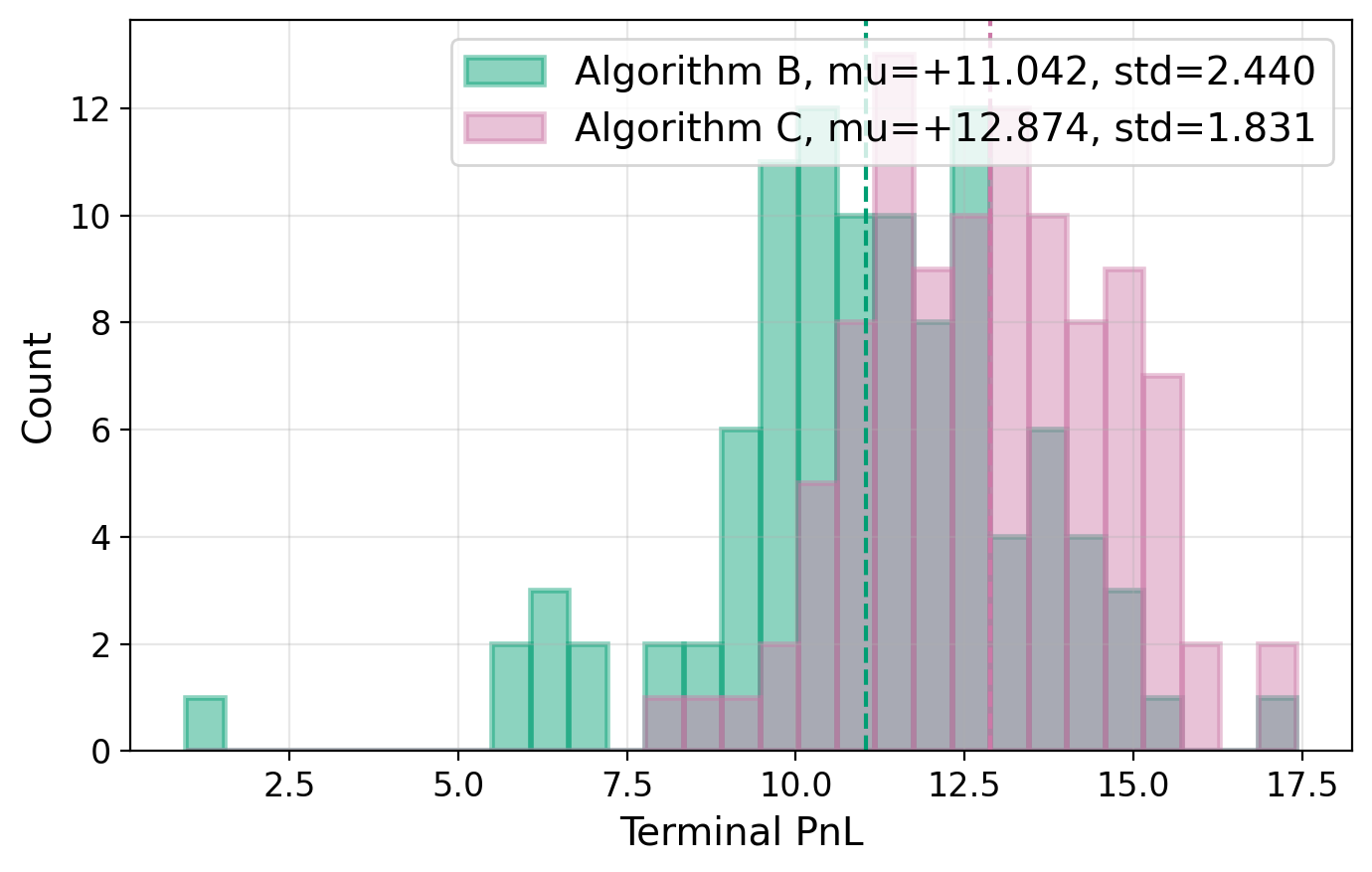}
    \caption{Correlated directional regimes.}
    \label{fig:bandit_correlated_regime_stress}
  \end{subfigure}
  \caption{
  Terminal-PnL distributions of Algorithm~B (green) and
  Algorithm~C (pink), evaluated over $100$ paired-seed episodes
  of $500\,000$ LOB events per panel. Relative to Algorithm~B,
  mean terminal PnL increases by approximately $14\%$ and
  $17\%$, respectively, while terminal-PnL SD falls to
  approximately $75\%$ of its Algorithm~B level in both tests.
  }
  \label{fig:bandit_combined}
\end{figure}

\paragraph{Stress-test evaluation.}
Figure~\ref{fig:bandit_combined} compares Algorithm~C with
frozen Algorithm~B under random regime persistence and
correlated directional flow. Across both stress tests, Algorithm~C increases mean terminal
PnL and reduces its dispersion.

With the same Algorithm~B initialization and fine-tuning budget,
adaptive scenario sampling and replacement outperformed uniform
sampling and random replacement. Mean terminal PnL increased by
$9.4\%$ under random regime persistence and $7.0\%$ under
correlated directional flow, with lower PnL dispersion in both
tests. This comparison uses one fine-tuning seed.

\section{Discussion and Conclusion}

Under stationary order flow, calibrated GLFT provides an
interpretable benchmark in the event-driven LOB, offering
a favorable trade-off between profitability and inventory
risk relative to the at-best strategy. The distributional
DQN achieves a better risk--return trade-off under the same
conditions. However, persistent directional flow substantially
deteriorates the performance of stationary controllers,
highlighting their vulnerability to changes in order-flow
dynamics.

Training under regime-switching flow (Algorithm~B), using
Bayesian estimates of directional imbalance and a
queue-exposure signal, restores positive mean PnL and
improves inventory control. Scenario-bandit fine-tuning
(Algorithm~C) further increases mean terminal PnL and
reduces its dispersion under random regime persistence
and correlated directional flow. These results highlight
the importance of accounting for nonstationary order flow
both in policy design and during training.

\paragraph{Limitations and future work.}
Our findings rely on a zero-intelligence Santa Fe LOB,
which does not capture strategic interactions among market
participants. The market maker submits unit-sized orders,
with at most one resting quote per side, while regime
durations are exponential and directional imbalance is
restricted to a fixed range. Further validation using
empirical LOB data and more realistic interactive
simulators would help assess the generality of these
results.

{\small
\bibliographystyle{unsrtnat}
\bibliography{references}
}

\appendix

\section{Bayesian Online Flow Filter}
\label{app:bayesian_flow_filter}

We apply Bayesian online change-point detection
\citep{adams2007bocpd,fearnhead2007online} to infer
directional order flow from market-order observations.
Throughout this appendix, $t$ indexes market orders
rather than all LOB events, with $y_t=1$ for a buy
market order and $y_t=0$ for a sell market order.
The filter does not observe the underlying regime
or its buy probability.

\subsection{Beta--Bernoulli Model}

Within each directional-flow segment, market-order
sides are conditionally independent given a fixed
buy probability $p$:
\[
  y_t\mid p\sim\mathrm{Bernoulli}(p),
  \qquad \iota=2p-1.
\]
Each new segment starts with a Beta prior
$p\sim\mathrm{Beta}(a_0,b_0)$.
After observing $B$ buys and $S$ sells within
the segment, conjugacy gives
\[
  p\mid B,S\sim\mathrm{Beta}(a,b),
  \qquad
  a=a_0+B,\quad b=b_0+S.
\]
The predictive probabilities for the next
market-order side are
\[
  \Prob(y=1\mid a,b)=\frac{a}{a+b},
  \qquad
  \Prob(y=0\mid a,b)=\frac{b}{a+b}.
\]
These probabilities enter the change-point
recursion below.

\subsection{Run-Length Filtering}

Let $\ell_t\geq 1$ denote the number of market orders
observed in the current segment, including $y_t$.
The filter maintains the run-length posterior
\[
  g_t(\ell)=\Prob(\ell_t=\ell\mid y_{1:t}),
\]
together with the conditional Beta posterior
\[
  p\mid \ell_t=\ell,y_{1:t}
  \sim
  \mathrm{Beta}\bigl(a_t^{(\ell)},b_t^{(\ell)}\bigr).
\]

For regime durations $L=\lceil X\rceil$, with
$X\sim\mathrm{Exp}(1/\tau_r)$, the probability
of a change point between consecutive market orders is
\[
  h=1-\exp(-1/\tau_r).
\]

For each run-length hypothesis $\ell$ at time $t-1$,
define the predictive probabilities
\[
  \mathcal E_t^{(\ell)}
  =
  \Prob\bigl(y_t\mid
  a_{t-1}^{(\ell)},b_{t-1}^{(\ell)}\bigr),
  \qquad
  \mathcal E_t^{(0)}
  =
  \Prob(y_t\mid a_0,b_0),
\]
using the Beta--Bernoulli predictive distribution
derived above.

If the current segment continues, its unnormalized
posterior weight and Beta parameters are updated as
\[
  \begin{aligned}
  \widetilde g_t(\ell+1)
  &=
  (1-h)g_{t-1}(\ell)\mathcal E_t^{(\ell)},\\
  a_t^{(\ell+1)}
  &=
  a_{t-1}^{(\ell)}+y_t,\\
  b_t^{(\ell+1)}
  &=
  b_{t-1}^{(\ell)}+1-y_t.
  \end{aligned}
\]

At a change point, $y_t$ is the first observation
of a new segment. Under the constant hazard,
the corresponding updates are
\[
  \begin{aligned}
  \widetilde g_t(1)
  &=
  h\mathcal E_t^{(0)},\\
  a_t^{(1)}
  &=
  a_0+y_t,\\
  b_t^{(1)}
  &=
  b_0+1-y_t.
  \end{aligned}
\]
We use $a_0=b_0=1$. Before normalization, we set
$\widetilde g_t(\ell)=0$ whenever $\ell>5\tau_r$ or
$\widetilde g_t(\ell)\leq10^{-10}$.
The retained weights are then normalized:
\[
  Z_t=\sum_{\ell\geq1}\widetilde g_t(\ell),
  \qquad
  g_t(\ell)=\frac{\widetilde g_t(\ell)}{Z_t}.
\]
The recursion applies for $t\geq2$, with
$g_1(1)=1$ and the initial Beta posterior
updated using $y_1$.

\subsection{Belief Summary Supplied to the Controller}

The belief $b_t$ of Eq.~\eqref{eq:phaseb_belief_summary}
summarizes the run-length posterior through
\[
  \begin{aligned}
  \widehat p_t
  &=
  \sum_{\ell\geq1}g_t(\ell)
  \frac{a_t^{(\ell)}}{a_t^{(\ell)}+b_t^{(\ell)}},\\
  \widehat{\iota}_t
  &=
  2\widehat p_t-1,\\
  \bar{\ell}_t
  &=
  \sum_{\ell\geq1}\ell\,g_t(\ell).
  \end{aligned}
\]
Here, $\widehat{\iota}_t$ estimates the directional
flow bias, while $\bar{\ell}_t$ is the posterior
expected length of the current segment.

Before entering the network, the expected run length
is transformed as
\[
  \widetilde{\ell}_t
  =
  \min\left\{
    1,\;
    \frac{\log(1+\bar{\ell}_t)}
         {\log(1+5\tau_r)}
  \right\},
\]
where $\tau_r$ is the assumed regime-duration scale.
The normalization uses the filter's truncation
threshold $5\tau_r$, beyond which the assumed
exponential regime duration has probability
$e^{-5}\approx0.7\%$.

\section{A Reduced GLFT Model Under Flow Asymmetry}
\label{app:glft_misspecification}

This appendix develops a reduced model of GLFT under
static directional-flow asymmetry. Quotes are computed
under symmetric order flow, while the buy-MO probability
$p\in(0,1)$ remains fixed throughout each episode.
We derive the stationary inventory distribution and
finite-horizon PnL approximation used in
Figure~\ref{fig:misspec_comparison}.

Throughout this appendix, $S_t$ is measured in ticks;
multiplication by $v_{\mathrm{tick}}=0.01$ converts price
changes to monetary units.

\subsection{GLFT Quotes and Effective Fill Intensities}

Let $q\in\{-Q,\ldots,Q\}$ denote inventory, with
$Q=q_{\max}$. From
Eqs.~\eqref{eq:glft_ask}--\eqref{eq:glft_bid},
the continuous GLFT quote distances, measured in ticks,
can be written as
\[
  \delta_a(q)=\bar\delta-c_{\mathrm{skew}}q,
  \qquad
  \delta_b(q)=\bar\delta+c_{\mathrm{skew}}q,
\]
where
\[
  \begin{aligned}
  \bar\delta
  &=
  \frac{1}{\gamma_{\mathrm{GLFT}}}
  \log\left(1+\frac{\gamma_{\mathrm{GLFT}}}{\kappa}\right)
  +\frac{c_{\mathrm{skew}}}{2},
  \\[4pt]
  c_{\mathrm{skew}}
  &=
  \sqrt{
    \frac{\sigma^2\gamma_{\mathrm{GLFT}}}{2\kappa A}
    \left(1+\frac{\gamma_{\mathrm{GLFT}}}{\kappa}\right)^{
    1+\kappa/\gamma_{\mathrm{GLFT}}}
  }.
  \end{aligned}
\]
These distances depend on inventory but not on
the realized flow asymmetry $p$.

To account approximately for differences between
the GLFT execution model and the FIFO simulator,
we introduce an effective intensity
$A_{\mathrm{eff}}=\eta A$, with
\[
  \eta
  =
  \frac{R_{\mathrm{empirical}}}
       {2Ae^{-\kappa\bar\delta}},
  \qquad
  R_{\mathrm{empirical}}
  =
  \frac{1}{E_{\mathrm{calib}}}
  \sum_{e=1}^{E_{\mathrm{calib}}}
  \frac{
    N_e^{a,\mathrm{fill}}+N_e^{b,\mathrm{fill}}
  }{T_e}.
\]
Here $R_{\mathrm{empirical}}$ is the average
two-sided execution rate under symmetric flow,
estimated from $E_{\mathrm{calib}}$ episodes
with durations $T_e$ measured in $\mathrm{u}$ and fill counts
$N_e^{a,\mathrm{fill}}$ and $N_e^{b,\mathrm{fill}}$.
The factor $\eta$ matches the GLFT zero-inventory
fill rate to the average rate observed under symmetric flow.

For a fixed buy-MO probability $p$, we approximate
the bid and ask fill intensities by

\begin{align}
  r_a(q;p)
  &=
  2p A_{\mathrm{eff}}
  e^{-\kappa\delta_a(q)},
  \label{eq:app_miss_ra}\\
  r_b(q;p)
  &=
  2(1-p) A_{\mathrm{eff}}
  e^{-\kappa\delta_b(q)}.
  \label{eq:app_miss_rb}
\end{align}
The factors $2p$ and $2(1-p)$ represent changes
in the distribution of market-order sides at fixed
total market-order intensity. At $p=1/2$, the
effective symmetric intensities are recovered.
Inventory limits impose
\[
  r_a(-Q;p)=0,
  \qquad
  r_b(Q;p)=0.
\]

The numerical analysis uses the calibrated
parameters of Eq.~\eqref{eq:glft_calib_values},
with $\sigma=0.30$ ticks / $\sqrt{\mathrm{u}}$,
$\gamma_{\mathrm{GLFT}}=10^{-6}$, $Q=8$,
and $\eta\simeq1.09$.
The corresponding quote coefficients are
$\bar\delta\simeq0.45$ ticks and
$c_{\mathrm{skew}}\simeq6.94\times10^{-4}$.
For subsequent PnL calculations, $T$
denotes a fixed horizon measured in $\mathrm{u}$,
and $v_{\mathrm{tick}}=0.01$ is the monetary
value of one tick.

\subsection{Stationary Inventory Distribution}

For fixed $p$, inventory follows a finite continuous-time
birth--death process on $\{-Q,\ldots,Q\}$. Bid fills
increase inventory at rate $r_b(q;p)$, while ask fills
decrease it at rate $r_a(q;p)$. Its stationary
distribution $\psi(q)$ satisfies the following detailed balance:
\[
  \psi(q)r_b(q;p)
  =
  \psi(q+1)r_a(q+1;p),
  \qquad q=-Q,\ldots,Q-1.
\]
Substituting Eqs.~\eqref{eq:app_miss_ra}--\eqref{eq:app_miss_rb}
gives
\[
  \frac{\psi(q+1)}{\psi(q)}
  =
  \frac{1-p}{p}
  \exp\{-\kappa c_{\mathrm{skew}}(2q+1)\}.
\]
Iterating this relation in both directions from $q=0$
and normalizing yields
\begin{equation}
  \psi(q)
  =
  \frac{1}{Z(\vartheta)}
  \exp\{-\kappa c_{\mathrm{skew}}q^2-\vartheta q\},
  \qquad
  Z(\vartheta)
  =
  \sum_{n=-Q}^{Q}
  \exp\{-\kappa c_{\mathrm{skew}}n^2-\vartheta n\},
  \label{eq:app_miss_pi}
\end{equation}
where
\[
  \vartheta=\log\left(\frac{p}{1-p}\right).
\]

\begin{figure}[htbp]
  \centering
  \begin{subfigure}[b]{0.50\linewidth}
    \centering
    \includegraphics[width=\linewidth]{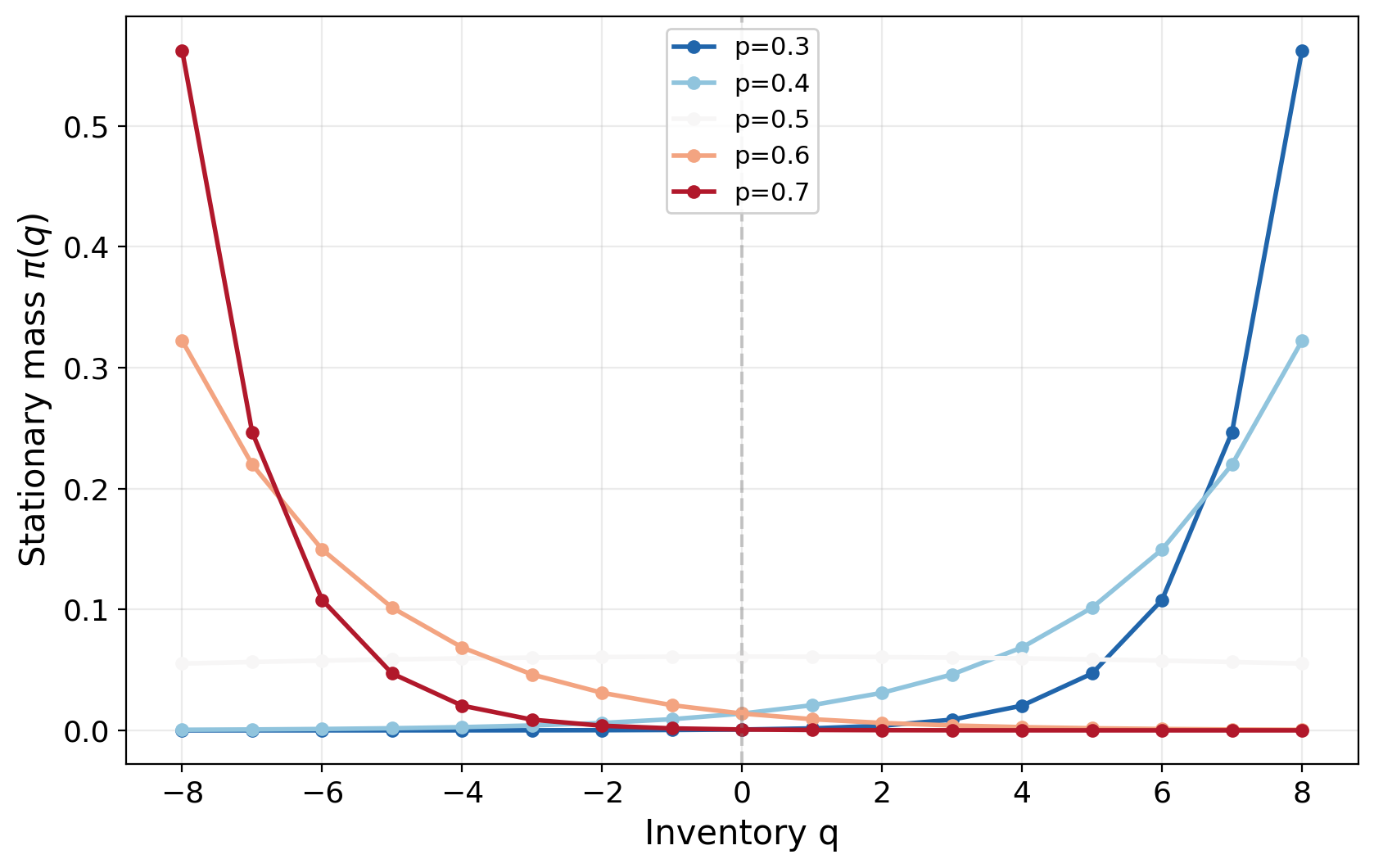}
    \caption{Selected $p$ values.}
    \label{fig:app_miss_pi_lines}
  \end{subfigure}%
  \begin{subfigure}[b]{0.50\linewidth}
    \centering
    \includegraphics[width=\linewidth]{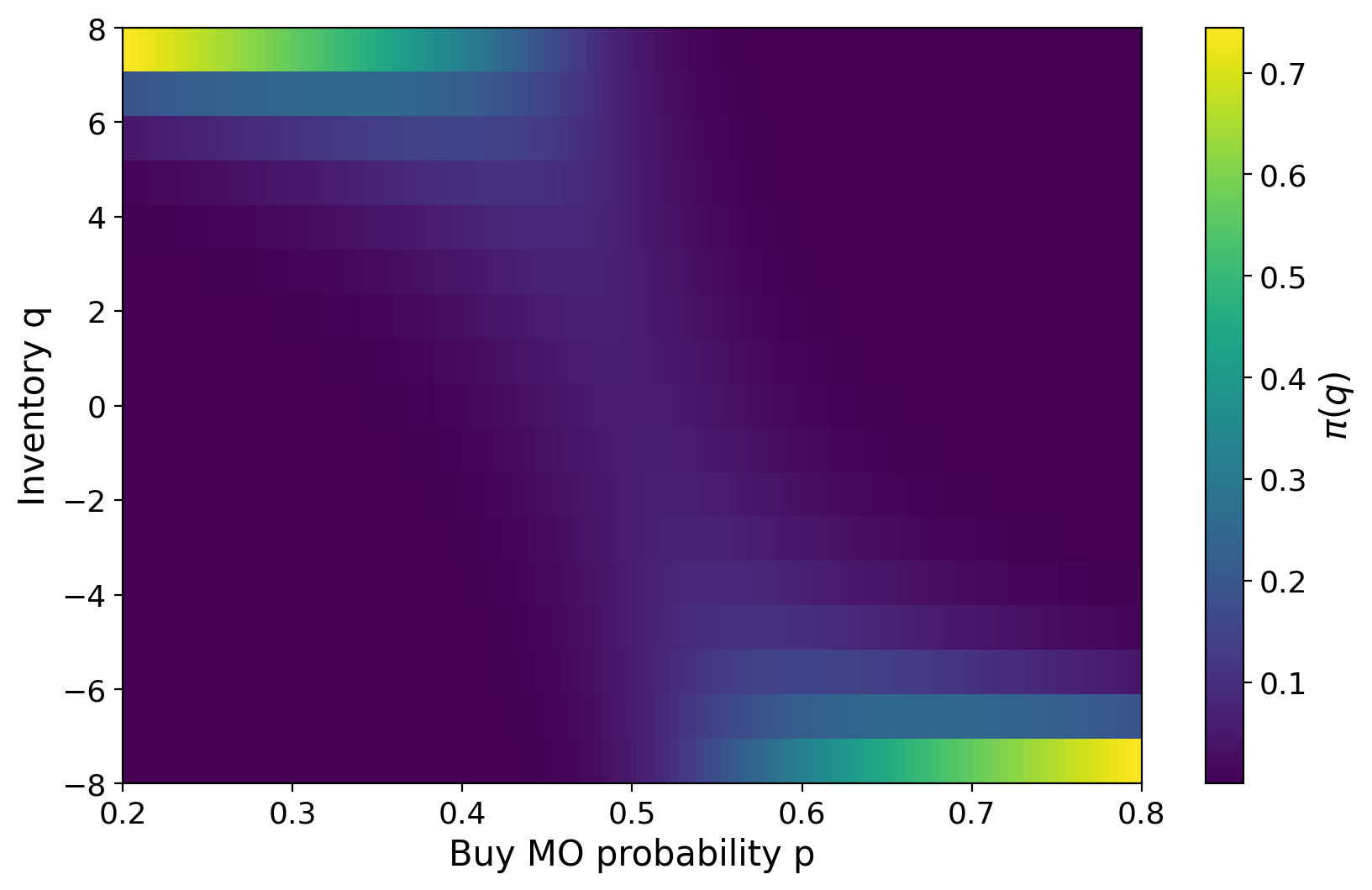}
    \caption{Full $p$ range.}
    \label{fig:app_miss_pi_heatmap}
  \end{subfigure}
  \caption{
  Stationary inventory distribution under GLFT
  misspecification for different buy-MO probabilities.
  }
  \label{fig:app_miss_stationary_law}
\end{figure}

The stationary distribution is a truncated discrete
Gaussian, with directional flow introducing a linear
tilt. Under symmetric flow ($p=1/2$), the distribution
is centered at zero. When $p>1/2$, more frequent ask
executions shift inventory toward negative values;
for $p<1/2$, the shift is reversed
(Figure~\ref{fig:app_miss_stationary_law}).

\subsection{Mean Inventory and Boundary Probabilities}

The stationary mean inventory follows directly
from Eq.~\eqref{eq:app_miss_pi}:
\begin{equation}
  \bar q(p)
  =
  \sum_{q=-Q}^{Q}q\,\psi(q)
  =
  -\frac{\partial}{\partial\vartheta}
  \log Z(\vartheta),
  \label{eq:app_miss_qbar}
\end{equation}
where $\vartheta=\log(p/(1-p))$.
Since $Z(\vartheta)$ is even, the mean inventory
is antisymmetric around $p=1/2$.
Buy-heavy flow shifts the distribution toward
short positions, while sell-heavy flow produces
the opposite effect.

\begin{figure}[htbp]
  \centering
  \begin{subfigure}[b]{0.50\linewidth}
    \centering
    \includegraphics[width=\linewidth]{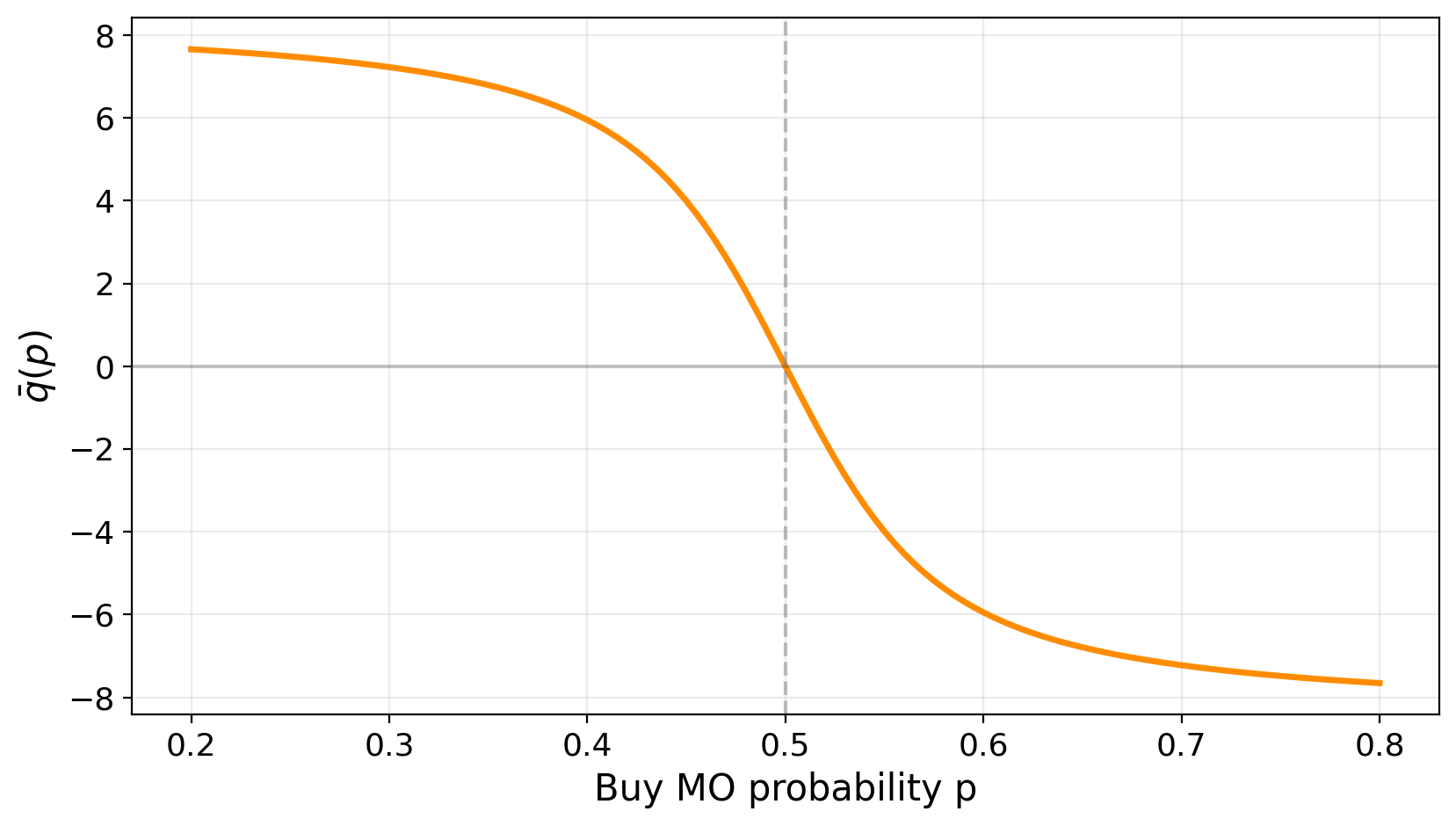}
    \caption{Mean inventory.}
    \label{fig:app_miss_qbar}
  \end{subfigure}%
  \begin{subfigure}[b]{0.50\linewidth}
    \centering
    \includegraphics[width=\linewidth]{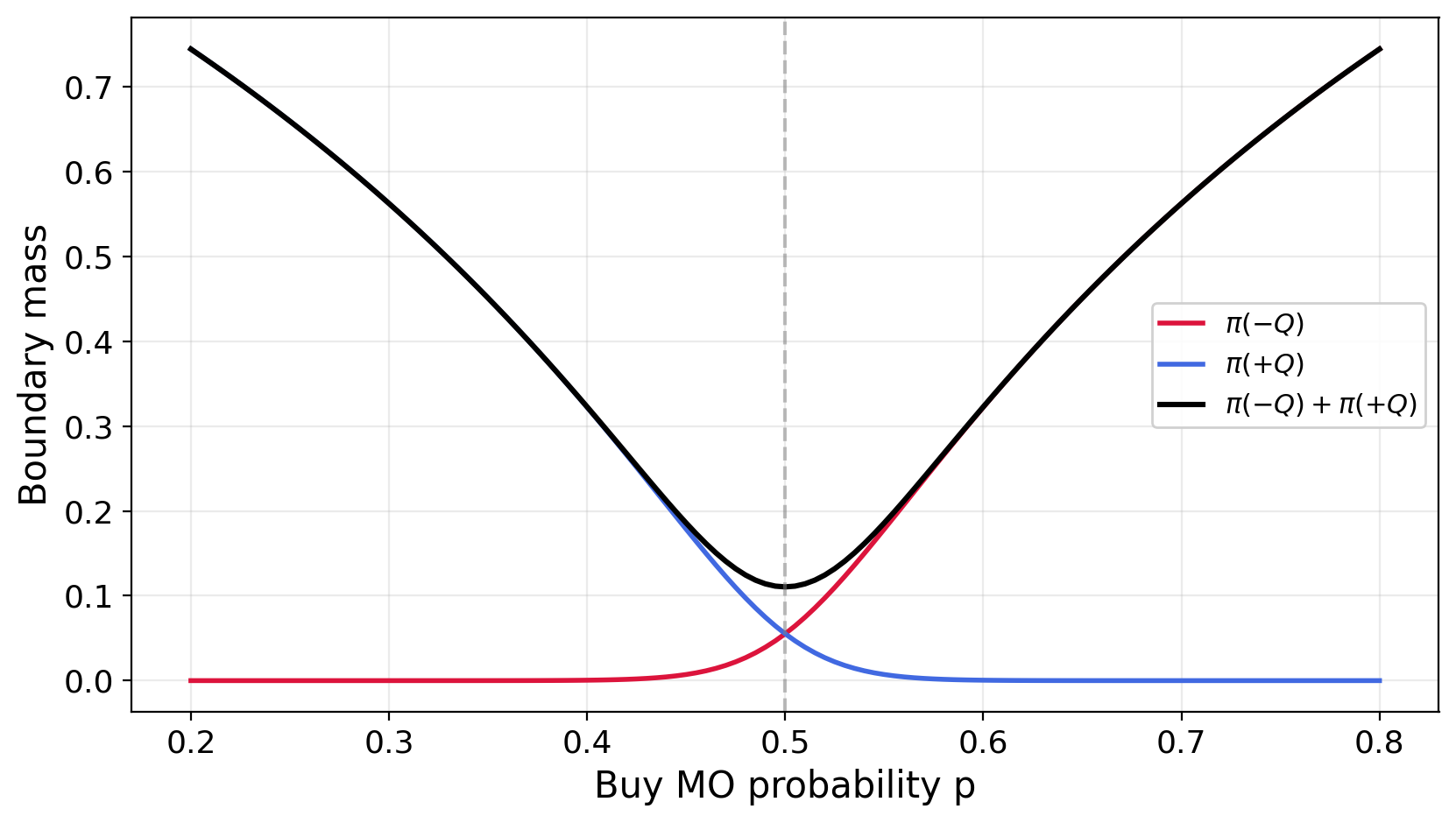}
    \caption{Boundary probability.}
    \label{fig:app_miss_boundary}
  \end{subfigure}
  \caption{
  Stationary inventory statistics under
  static directional-flow asymmetry.
  }
  \label{fig:app_miss_inventory_diagnostics}
\end{figure}

As flow asymmetry increases, the stationary
probability of occupying an inventory limit,
$\psi(-Q)+\psi(Q)$, also increases
(Figure~\ref{fig:app_miss_inventory_diagnostics}).
At these limits, quoting on the corresponding
side is suppressed.

\subsection{Expected Spread Capture}

The expected spread-capture rate at inventory $q$ is
\[
  g(q;p)
  =
  r_a(q;p)\delta_a(q)+r_b(q;p)\delta_b(q).
\]
For the finite-horizon analysis, inventory starts at
$q_0=0$. Let
$\rho_t(q;p)=\Prob(q_t=q\mid q_0=0)$ denote its
distribution at time $t$, obtained from the
birth--death generator through a matrix exponential.
Define the time-averaged distribution
\[
  \bar\rho_T(q;p)
  =
  \frac{1}{T}\int_0^T\rho_t(q;p)\,\dd t.
\]
Expected spread capture over horizon $T$ is then
\begin{equation}
  \E[\mathrm{PnL}_{\mathrm{spread}}](p)
  =
  v_{\mathrm{tick}}T
  \sum_{q=-Q}^{Q}\bar\rho_T(q;p)g(q;p).
  \label{eq:app_miss_spread_pnl}
\end{equation}
For stationary initialization, $\bar\rho_T(q;p)$
is replaced by $\psi(q)$.

Under the GLFT martingale mid-price assumption,
$\dd S_t=\sigma\,\dd W_t$, bounded inventory implies
\[
  \E\left[\int_0^T q_{t-}\,\dd S_t\right]=0.
\]
Thus, Eq.~\eqref{eq:app_miss_spread_pnl} also gives
the expected total PnL in the absence of
directional price drift.
  \subsection{Drift-Corrected Curve}

To capture the adverse mark-to-market effect of persistent
directional flow, we consider the reduced-form price model
\[
  \dd S_t
  =
  v_0(2p-1)\,\dd t+\sigma\,\dd W_t,
\]
where $2p-1$ is the signed order-flow imbalance and
$v_0>0$ converts this imbalance into a price drift.
This correction is not part of the original GLFT
martingale benchmark; setting $v_0=0$ recovers
Eq.~\eqref{eq:app_miss_spread_pnl}.

We calibrate $v_0$ from simulator runs with fixed
buy-MO probability $p_j$. For episode $e$ at level $p_j$,
with duration $T_{j,e}$ measured in $\mathrm{u}$,
the realized mid-price drift is
\[
  \widehat b_{j,e}
  =
  \frac{S_T^{(j,e)}-S_0^{(j,e)}}{T_{j,e}},
\]
and its average across episodes is
\[
  \bar b_j
  =
  \frac{1}{E_j}\sum_{e=1}^{E_j}\widehat b_{j,e}.
\]
We then estimate $v_0$ by the zero-intercept
least-squares slope of $\bar b_j$ on $2p_j-1$:
\[
  \widehat v_0
  =
  \frac{\sum_j (2p_j-1)\bar b_j}
       {\sum_j (2p_j-1)^2}.
\]
For the baseline simulator, this gives
$\widehat v_0\approx 0.05$ ticks/$\mathrm{u}$.

The finite-horizon PnL curves use a fixed reference horizon
$T$, measured from a symmetric simulator run with the same
event horizon as the analytical comparison. In the plots,
$T\approx 844\,\mathrm{u}$, approximating the
random duration of the $5{,}000$-event FIFO episodes.

\begin{figure}[!htbp]
  \centering
  \begin{subfigure}[b]{0.43\linewidth}
    \centering
    \includegraphics[width=\linewidth]{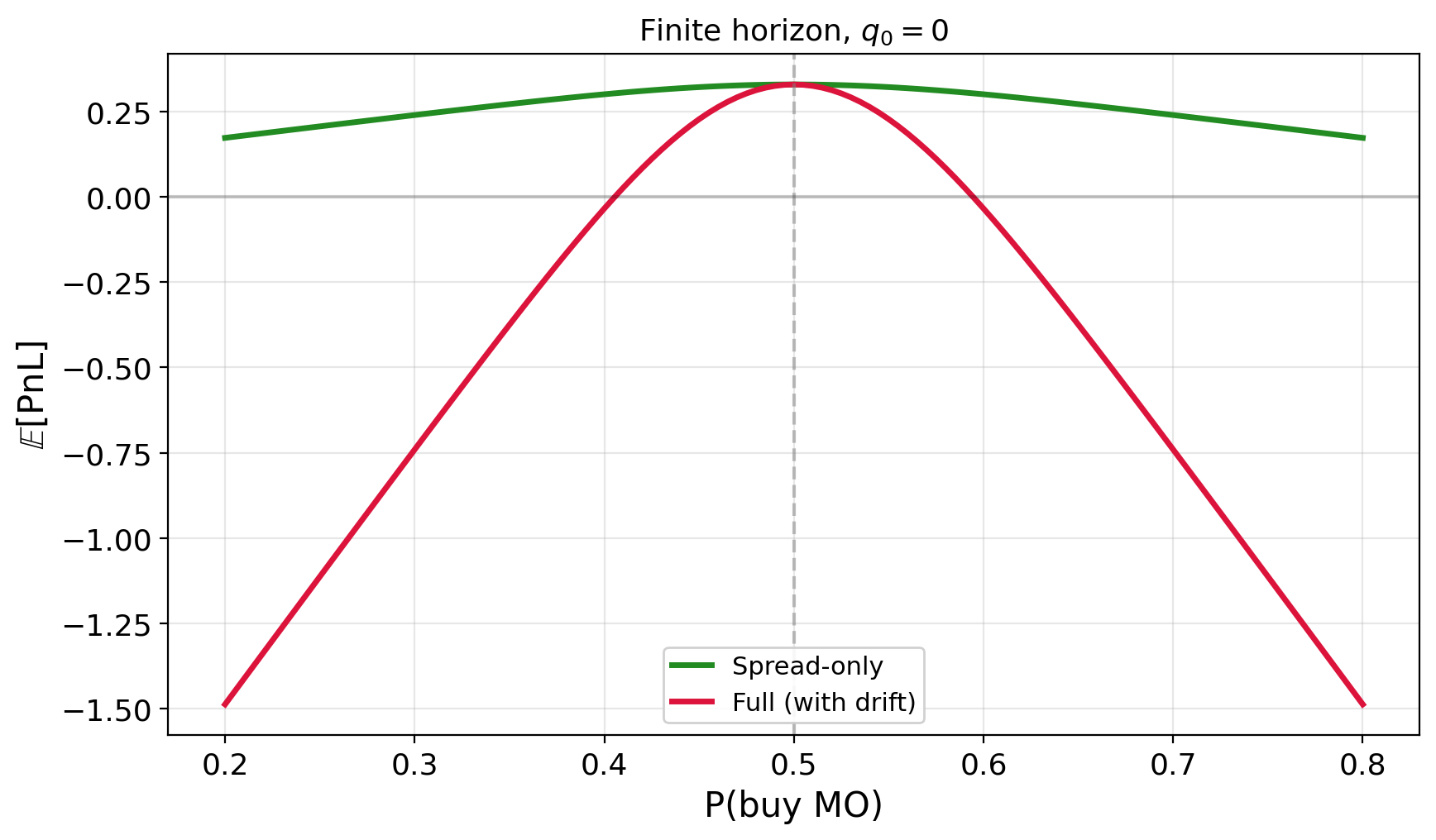}
    \caption{Expected PnL.}
    \label{fig:app_miss_pnl_curves}
  \end{subfigure}
  \hfill
  \begin{subfigure}[b]{0.43\linewidth}
    \centering
    \includegraphics[width=\linewidth]{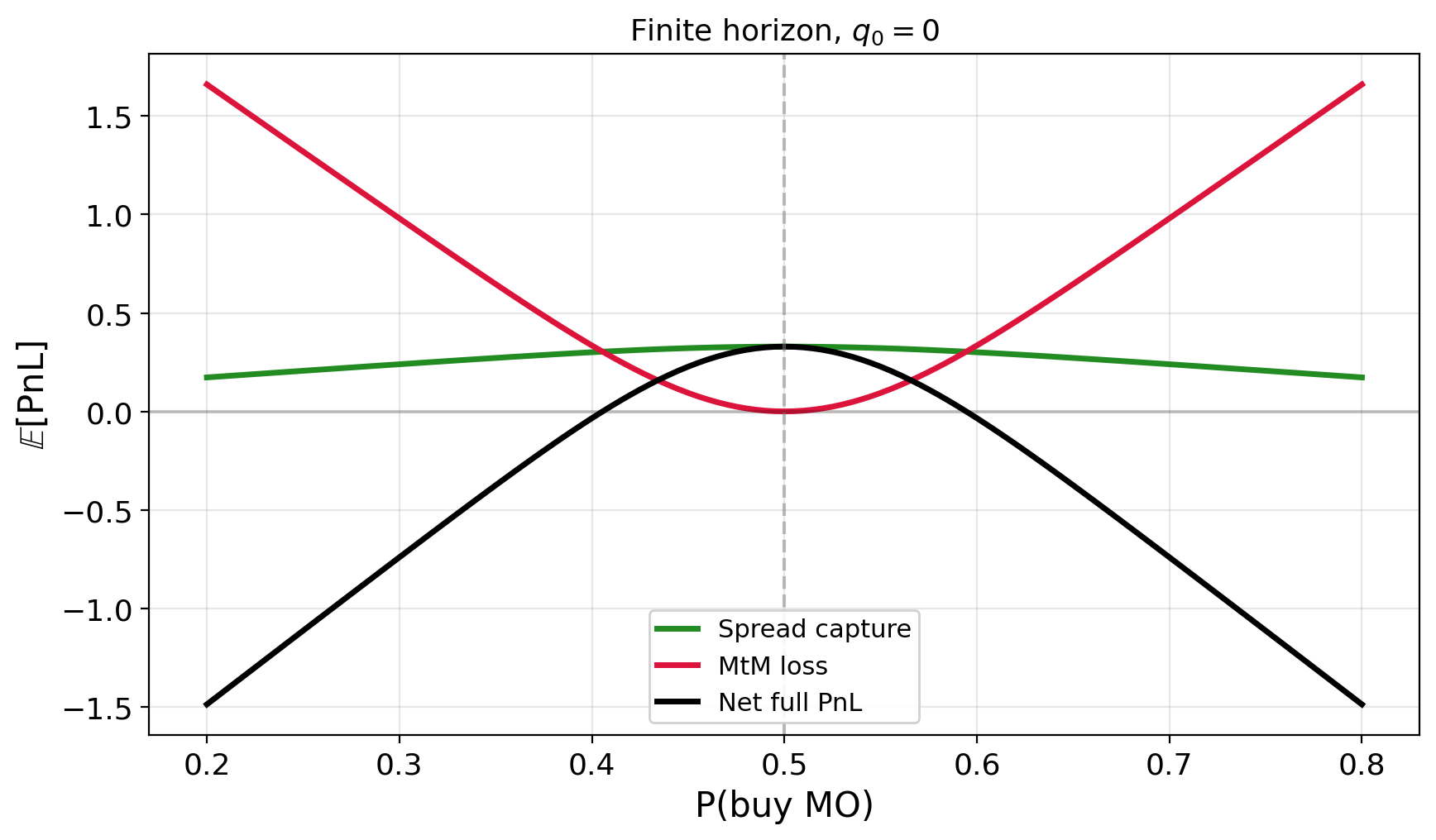}
    \caption{PnL decomposition.}
    \label{fig:app_miss_pnl_decomp}
  \end{subfigure}
  \caption{
  Finite-horizon PnL under static flow asymmetry,
  with $q_0=0$ and $T\approx844\,\mathrm{u}$.
  }
  \label{fig:app_miss_pnl}
\end{figure}

Under this reduced-form drift model,
\[
  \E\left[\int_0^T q_{t-}\,\dd S_t\right]
  =
  v_0(2p-1)T
  \sum_{q=-Q}^{Q} q\,\bar\rho_T(q;p),
\]
so the drift-corrected expected PnL is
\begin{equation}
  \E[\mathrm{PnL}_{\mathrm{full}}](p)
  =
  \E[\mathrm{PnL}_{\mathrm{spread}}](p)
  +
  v_{\mathrm{tick}}v_0(2p-1)T
  \sum_{q=-Q}^{Q} q\,\bar\rho_T(q;p).
\end{equation}
Because the inventory averaged over time has the opposite sign
as $2p-1$, this correction is adverse on both sides of
$p=1/2$ (Figure~\ref{fig:app_miss_pnl}).

\end{document}